\documentclass{aa}  
\usepackage[dvipsnames]{xcolor} 
\usepackage[para,online,flushleft]{threeparttable}
\usepackage{siunitx}
\usepackage{CJKutf8}
\usepackage{enumitem}
\usepackage{afterpage}
\usepackage{graphicx}
\usepackage{txfonts}
\usepackage{hyperref}
 \hypersetup{
     colorlinks=true,
     linkcolor=purple,
     citecolor=blue, 
     }
     
\begin{document} 

\title{Gravitational signal propagation in the Double Pulsar \\studied with the MeerKAT telescope}

\titlerunning{Gravitational signal propagation in the Double Pulsar}

\author{H.~Hu (胡奂晨)
  \inst{1}
  \and M.~Kramer \inst{1} \fnmsep\inst{2}
  \and D.~J.~Champion \inst{1}
  \and N.~Wex \inst{1}
  \and A.~Parthasarathy \inst{1}
  \and T.~T.~Pennucci \inst{3}
  \and N.~K.~Porayko \inst{1}
  \and W.~van~Straten \inst{4}
  \and V.~Venkatraman~Krishnan \inst{1}
  \and M.~Burgay \inst{5}
  \and P.~C.~C.~Freire \inst{1}
  \and R.~N.~Manchester \inst{6}
  \and A.~Possenti \inst{5}
  \and I.~H.~Stairs \inst{7}
  \and M.~Bailes \inst{8}\fnmsep\inst{9}
  \and S.~Buchner \inst{10}
  \and A.~D.~Cameron \inst{8}\fnmsep\inst{9}
  \and F.~Camilo \inst{10}
  \and M.~Serylak \inst{11}\fnmsep\inst{12}
}
\authorrunning{H. Hu et al.}
\institute{Max-Planck-Institut f\"ur Radioastronomie, Auf dem H\"ugel 69, 53121 Bonn, Germany \\
\email{huhu@mpifr-bonn.mpg.de}
\and Jodrell Bank Centre for Astrophysics, The University of Manchester, Oxford Road, Manchester M13 9PL, United Kingdom
\and Institute of Physics, E\"{o}tv\"{o}s Lor\'{a}nd University, P\'{a}zm\'{a}ny P.s. 1/A, 1117 Budapest, Hungary
\and Institute for Radio Astronomy \& Space Research, Auckland University of Technology, Private Bag 92006, Auckland 1142, New Zealand
\and INAF-Osservatorio Astronomico di Cagliari, via della Scienza 5, 09047, Selargius, Italy
\and Australia Telescope National Facility, CSIRO Space and Astronomy, P.O. Box 76, Epping NSW 1710, Australia
\and Dept. of Physics and Astronomy, University of British Columbia, 6224 Agricultural Road, Vancouver, BC V6T 1Z1 Canada
\and Centre for Astrophysics and Supercomputing, Swinburne University of Technology, PO Box 218, Hawthorn, VIC 3122, Australia
\and ARC Centre of Excellence for Gravitational Wave Discovery (OzGrav), Mail H29, Swinburne University of Technology, PO Box 218, Hawthorn, VIC 3122, Australia
\and  South African Radio Astronomy Observatory, Cape Town 7925, South Africa
\and SKA Observatory, Jodrell Bank, Lower Withington, Macclesfield, SK11 9FT, United Kingdom
\and Department of Physics and Astronomy, University of the Western Cape, Bellville, Cape Town, 7535, South Africa
}

\date{Received MM dd, yyyy; accepted MM dd, yyyy}

\abstract{The Double Pulsar,  PSR~J0737$-$3039A/B, has offered a wealth of gravitational experiments in the strong-field regime, all of which general relativity has passed with flying colours. In particular, 
among current gravity experiments that test photon propagation, the Double Pulsar probes the strongest spacetime curvature.
Observations with MeerKAT and, in future, the Square Kilometre Array (SKA) can greatly improve the accuracy of current tests and facilitate tests of next-to-leading-order (NLO) contributions in both orbital motion and signal propagation.We present our timing analysis of new observations of PSR~J0737$-$3039A, made using the MeerKAT telescope over the last three years.
The increased timing precision offered by MeerKAT yields a 2 times better measurement of Shapiro delay parameter $s$ and improved mass measurements compared to previous studies.
In addition, our results provide an independent confirmation of the NLO signal propagation effects and already surpass the previous measurement from 16-yr data by a factor of 1.65.
These effects include the retardation effect due to the movement of the companion and the deflection of the signal by the gravitational field of the companion. 
We also investigate novel effects which are expected. For instance, we search for potential profile variations near superior conjunctions caused by shifts of the line-of-sight due to latitudinal signal deflection and find insignificant evidence with our current data. 
With simulations, we find that the latitudinal deflection delay is unlikely to be measured with timing because of its correlation with Shapiro delay. 
Furthermore, although it is currently not possible to detect the expected lensing correction to the Shapiro delay, our simulations suggest that this effect may be measured with the full SKA.
Finally, we provide an improved analytical description for the signal propagation in the Double Pulsar system that meets the timing precision expected from future instruments such as the full SKA.
}

\begin{CJK*}{UTF8}{gkai}
\maketitle
\end{CJK*}

\section{Introduction}
The Double Pulsar PSR~J0737$-$3039A/B is a rich laboratory for strong-field gravity experiments. The system consists of a 23-ms recycled pulsar (``A'') and a 2.8-s ``normal'' pulsar (``B'') in a nearly edge-on and slightly eccentric 2.45-hr orbit \citep{Burgay+2003,Lyne+2004}. Various relativistic effects have been precisely measured in previous works \citep{Kramer+2006,Kramer+2021DP}, including periastron precession, time dilation (gravitational redshift and second-order Doppler effect), Shapiro delay due to light propagation in the curved spacetime of the companion, and the orbital period decay which currently provides the most precise test of quadrupolar gravitational wave predicted by general relativity (GR). In addition, the relativistic spin precession of B was measured by \citet{Breton+2008} and the relativistic deformation of the orbit was newly detected in this system \citep{Kramer+2021DP}. All these make it a still unique system for gravity experiments.

Comparing with other gravity experiments, the Shapiro delay measured in the Double Pulsar probes the strongest spacetime curvature ($\sim 10^{-21}\,\mathrm{cm^{-2}}$) in a precision experiment with photons, i.e. the interaction between gravitational and electromagnetic fields \citep{WK2020}. In addition, with 16~yr of data, \citet{Kramer+2021DP} were able for the first time to measure higher-order effects of signal propagation in the strong gravitational field of a neutron star, which are currently not accessible via any other method. 
These include retardation effect due to the movement of the companion (B) and aberrational light deflection by the gravitation of the companion. The latter confirms the prograde rotation of A, which is consistent with the results measured by \citet{Pol_2018} using the emission properties of B and what is expected from binary evolution models. 

In this work, we present observations of PSR~J0737$-$3039A with the MeerKAT telescope, a precursor for the Square Kilometre Array (SKA) at mid-frequency range. Thanks to its location in the Southern Hemisphere, it permits a timing precision more than two times better than that of the Green Bank Telescope for this pulsar \citep{Bailes+2020, Kramer+2021Relbin}. This superior precision enables an independent  and improved measurement of signal propagation effects within a very short time span. We also investigate effects that have been expected but not been studied in detail before. These include potential profile variations due to latitudinal deflection, the detectability of latitudinal deflection delay, and the prospects of measuring the effect of lensing on the propagation time separately.

This paper is organised as follows. Section~\ref{sec:obs} describes the MeerKAT observations on the Double Pulsar and data processing. In Section~\ref{sec:NLO}, we introduce the concepts of gravitational signal propagation effects, including higher-order contributions to the Shapiro and aberration delay. 
The timing results and mass measurements are presented in Section~\ref{sec:timing}. We then provide an in-depth study on the higher-order signal propagation effects in Section~\ref{sec:test_NLO}, with focus on latitudinal deflection and lensing. In addition, we provide an improved analytical description for the signal propagation in the Double Pulsar. Finally, we discuss the results and future prospects in Section~\ref{sec:dis}.

\section{Observations and data processing}
\label{sec:obs}

\subsection{MeerKAT observations}
The observations presented in this paper come from the MeerKAT telescope as part of the MeerTIME project \citep{Bailes+2020}, which performs timing of known pulsars with various scientific themes.
Observations on PSR~J0737$-$3039A are conducted under the Relativistic Binary theme \citep[RelBin,][]{Kramer+2021Relbin}, which focuses on testing the relativistic effects in binary pulsars to achieve measurements of neutron star masses and tests of theories of gravity.
MeerTime observations are generally recorded using the Pulsar Timing User Supplied Equipment (PTUSE) signal processor. This processor receives channelised tied-array beamformed voltages from the correlator-beamformer engine of the MeerKAT observing system and is capable of producing coherently de-dispersed full-Stokes data in both filterbank (search) mode and fold (timing) mode, where the data are folded at the topocentric period of the pulsar. Details on pulsar observing set up with MeerTime are explained by \citet{Bailes+2020}.

PSR~J0737$-$3039A is regularly observed with a typical cadence of one month and duration of 3~hr. As the orbital period of this pulsar is $\sim2.45$~hr, the observations are scheduled to start shortly before an eclipse and finish after the second eclipse, in order to observe the eclipses twice in one observing session. The session is typically composed of a 30-min observation with fold mode and search mode in parallel, followed by a 2-hr fold-mode observation and another 30-min fold-search dual-mode observation. This specific arrangement is designed to maximise our sensitivity in detecting signal propagation effects, as well as in studying the magnetosphere of pulsar B (Lower et al. in prep.). Observations are performed with two receivers: The L-band receiver that covers the frequency range 856--1712~MHz and the UHF receiver that covers the frequency range 544--1088~MHz, both with 1024 channels. The data presented here starts in March 2019 and runs to May 2022. For the analysis in this paper, we use 29 full-orbit timing observations and 62 search-mode eclipse data sets, which amounts to a total of $\sim 87$~hr.

{\renewcommand{\arraystretch}{1.2}
\begin{table*}[t]
    \caption{Information on MeerKAT observation and data set$^a$ for PSR~J0737$-$3039A.}
    \centering
    \begin{threeparttable}
    \begin{tabular}{ccccccc}
    \hline \hline
    Receiver &  Centre Frequency & Bandwidth  & Number of & Number of & Time span  & Number of  \\
    & (MHz) & (MHz) & channels$^b$ & sub-bands & (MJD) & TOAs
    \\ \hline
    L-band   & 1283.582 & 775.75 & 928 & 16 & 58568 - 59721 & 83930 \\
    UHF-band &  815.734 & 493    & 928 & 32 & 58936 - 59663 & 137451 \\\hline
    \end{tabular}
    \begin{tablenotes}
    $^a$ Information presented here are for the trimmed data set, see Section~\ref{sec:timing_par}. \\
    $^b$ Effective usable channels.\\
    \end{tablenotes}
    \end{threeparttable}
    \label{tab:data}
\end{table*}
}
\subsection{Timing data reduction}
\label{sec:data_reduce}
The raw 8~s-folded timing data from the PTUSE machines are processed with the \textsc{meerpipe} data reduction pipeline. \textsc{meerpipe} carries out radio frequency interference (RFI) removal using a modified \textsc{coastguard} algorithm \citep{Lazarus+2016}, followed by flux and polarisation calibration. Details on polarisation and flux calibration are described in \citet{Serylak+2021} and \citet{Spiewak+2022}, respectively.

After processing with \textsc{meerpipe}, the calibrated data products are reduced using pulsar software package \textsc{psrchive}\footnote{\url{http://psrchive.sourceforge.net/}} \citep{Hotan+2004}. 
We first correct for the rotation measure (RM) with the value measured in \citet{Kramer+2021Relbin}, i.e. RM=$120.84\, \mathrm{rad\,m^{-2}}$. As the L-band observations between March 2019 and February 2020 are restricted to 928 frequency channels (dropping 48 channels each from the top and bottom bands), to maintain consistency throughout the analysis, we reduce the later L-band data to the same frequency channels. We treat the UHF-band data in the same way, as the roll-off adversely affects sensitivity of the top and bottom bands.

For this system, a complete timing model is only available in the pulsar timing software \textsc{tempo}\footnote{\url{http://tempo.sourceforge.net/}} \citep[][more details will be given in Section~\ref{sec:timing}]{TEMPO:2015}. Therefore, to fold the data more accurately, all data are supplied with a \texttt{polyco}-format ephemeris with the values measured in \citet{Kramer+2021DP}. Since the Double Pulsar rapidly changes its orbital phase, the time span (TSPAN) of a predicted pulse phase solution has to be as small as possible to keep a good precision\footnote{Our analysis suggests that the choice of TSPAN has a significant impact on the Shapiro parameters, a larger TSPAN leads to a large deviation from the expected values.}. With the \textsc{psrchive} version 2022-01-14, we set TSPAN to the minimum possible value which is 3~min. 

A known data processing issue with this pulsar is that the pulsar moves rapidly to a different orbital phase during the dispersion delay time, hence the pulses received at the same time at different frequencies correspond to different orbital phases, therefore can not be folded with the same phase prediction. 
If not properly accounted for, this folding issue will cause frequency-dependent orbital smearing. Standard pulsar software like \textsc{psrchive} does not take this effect fully into consideration even with frequency-resolved \textsc{tempo2} predictor\footnote{This issue is going to be addressed in \textsc{psrchive}~2.0 under development.}.
To avoid this issue, we first de-disperse the total intensity data so that all frequencies correspond to the same orbital phase, then average the data first in frequency and then in time\footnote{The order of processing matters. If reversed, the pulse phase appears to be different and phase offsets may be introduced.}. Because of the profile frequency evolution and scintillation effects, data are sub-banded in frequency, with 32 sub-bands for the UHF-band and 16 sub-bands for the L-band.

As for the time-averaging, the integration time needs to be short enough in order to properly resolve the Shapiro delay and the next-to-leading-order (NLO) signal propagation contributions ($q_\mathrm{NLO}$, see Section~\ref{sec:NLO}), which are largest at superior conjunction. 
We perform a simulation to test the measurability of these NLO contributions with different integration time. The results show that a good measurement of Shapiro delay and $q_\mathrm{NLO}$ can be achieved if the integration time is $\lessapprox 32$~s, but it becomes significantly worse if the integration time is longer than 1~min for $q_\mathrm{NLO}$ and 2~min for Shapiro delay. Therefore, we average all data with 32~s integration time, consistent with the analysis by \citet{Kramer+2021DP}. 
After frequency and time averaging, data  are re-dispersed to allow measurement of dispersion measure (DM) in the timing analysis. 
\begin{figure}[t]
    \vspace{-10pt}
    \centering
    \includegraphics[width=0.9\columnwidth]{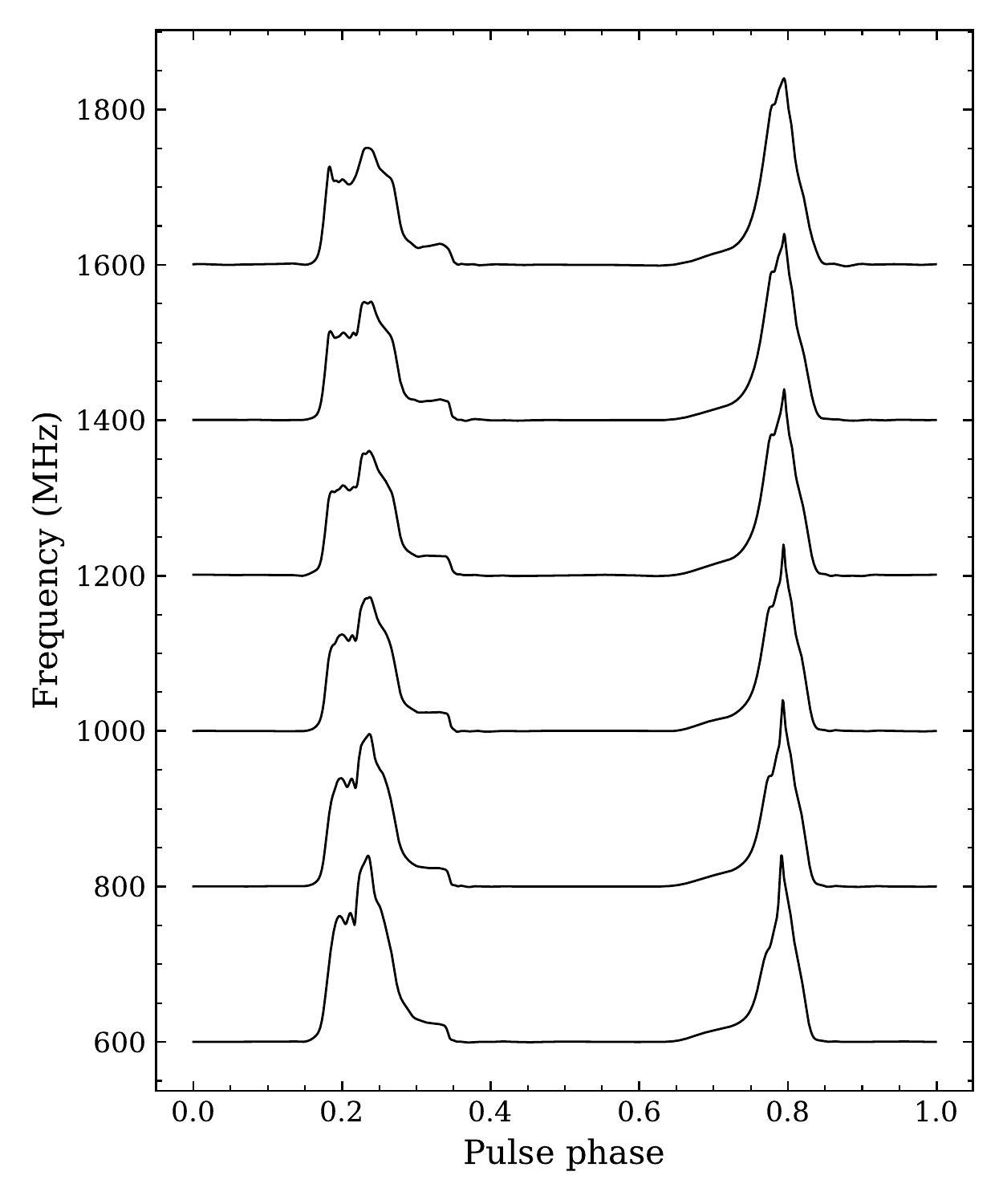}
    \caption{Pulse profile of PSR J0737$-$3039A observed at multiple frequencies with the MeerKAT UHF- and L-band receivers.}
    \label{fig:profile}
\end{figure}

\subsection{Wide-band templates and TOA extraction}
\label{sec:wideband}
\begin{figure}[t]
    \vspace{-10pt}
    \centering
    \includegraphics[width=\columnwidth]{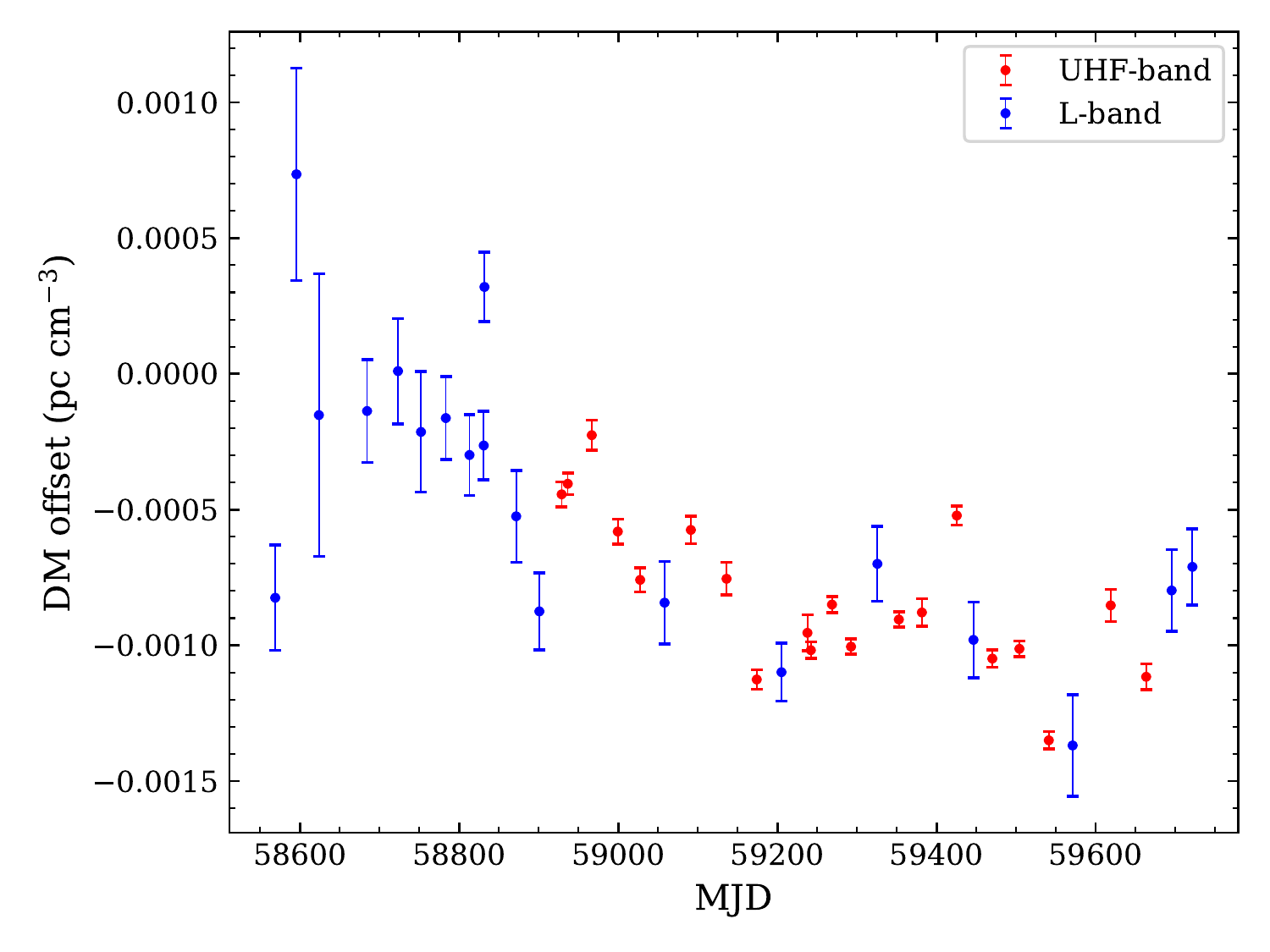}
    \vspace{-15pt}
    \caption{DM measurement per observing epoch relative to the reference value $48.913 \mathrm{\,pc\,cm^{-3}}$ (see Section~\ref{sec:dm}).}
    \label{fig:dm}
\end{figure}

Wide-band observations like MeerTIME can suffer significant profile evolution in frequency, hence the traditional 1D template is not favoured. To best determine the pulse time-of-arrivals (TOAs) at multiple frequencies, we employ frequency-dependent 2D templates. With this technique, DM measurements are dependent on the DM value used to align (de-disperse) the 2D template.
Due to the correlation between DM and profile evolution, DM measurements are to some extend frequency dependent, which can lead to a DM offset between L-band and UHF-band data. This could potentially be solved with a simultaneous observation with L-band and UHF-band receivers, which is missing in our case. Therefore, to avoid this problem, we choose a bright observation from each band for making 2D templates, and measure DM using data from their overlapping frequencies. Then, we use these DM values to de-disperse the corresponding full-bandwidth data.
This minimises the DM offset between L-band and UHF-band data which can be seen in Fig.~\ref{fig:dm}. 
These data are then sub-banded and averaged in time. Finally, by smoothing the profiles with \texttt{psrsmooth}/\textsc{psrchive}, we obtain 2D templates, with 16 sub-bands at L-band and 32 sub-bands at UHF-band. These templates are then used to measure frequency-resolved TOAs by cross-correlating with the reduced data using \texttt{pat}/\textsc{psrchive}. The pulse profile of PSR~J0737$-$3039A at multiple frequencies is shown in Fig.~\ref{fig:profile}.
More information on the observing systems and data sets is given in Table~\ref{tab:data}.

\subsection{DM variation}
\label{sec:dm}
The wide-band observation and high precision of MeerKAT telescope make it possible to obtain an accurate DM measurement on a per-epoch basis so as to minimise the influence of DM noise in the data. To do so, we fit for only DM and spin frequency $\nu$ for each observing epoch using 4-min TOAs, and keep the other parameters fixed. The DM measurements are shown in Fig.~\ref{fig:dm}. Following \citet{Kramer+2021DP}, we use a modified version of \textsc{tempo} for our timing analysis, which corrects dispersive delays for each TOA based on the exact DM measurement of that epoch.

One should note that our data set does not show the apparent DM variation as a function of orbital phase as was seen in \citet{Ransom+2004}. It had been demonstrated that this effect occurs due to an unaccounted Doppler shift of the observational frequency as the pulsar moves in a binary system\footnote{\url{https://arxiv.org/e-print/astro-ph/0406321v2}}, this will be revisited by Hu, Porayko et al. (in prep.). 
More thorough investigation of this effect as well as the frequency-dependent orbital smearing (see Section~\ref{sec:data_reduce}) is ongoing and will be presented in detail in the future publication.


\section{Signal propagation effects at superior conjunction}
\label{sec:NLO}

In this section, we recapitulate the necessary concepts of signal propagation effects in the Double Pulsar, including the NLO contributions in the Shapiro delay and aberration delay, which were described in greater detail in \citet{Kramer+2021DP}.

Being a nearly edge-on binary system (i.e. $i\sim 90\si{\degree}$), the curved spacetime of the companion star (pulsar B) has a significant effect on the propagation of the pulsar's signal. To leading-order this is the well-known Shapiro delay \citep{Shapiro1964}, which is expressed in the following form for binary pulsars \citep{BT1976,DD86}:
\begin{align}
    \Delta_\mathrm{S}^\mathrm{(LO)} = &-2r \ln{\Lambda_u} \,,\label{eq:s_lo}\\ 
   \Lambda_u = &1-e_\mathrm{T} \cos{u} - s\, \Big[\sin{\omega}\,(\cos{u}-e_\mathrm{T}) \nonumber\\
   &+ (1-e_\mathrm{T}^2)^{1/2} \cos{\omega}\sin{u}\Big] \,.\label{eq:shapiro}
\end{align}
Here, $u$ denotes the eccentric anomaly (from Kepler's equation with eccentricity $e_\mathrm{T}$), and $\omega$ denotes the longitude of periastron measured from the ascending node. The time eccentricity $e_\mathrm{T}$ corresponds to the eccentricity parameter in the Damour-Deruelle (DD) timing model \citep{DD86} that can be fitted in pulsar timing software \textsc{tempo} or \textsc{tempo2} \citep{tempo2}.
The two post-Keplerian (PK) parameters $r$ and $s$ represent the \emph{range} and \emph{shape} of Shapiro delay, respectively. 
The shape parameter is generally identified with the sine of the orbital inclination $i$ as $s\equiv\sin{i}$, whereas the range parameter is linked to the mass of the companion $m_\mathrm{B}$, which in GR follows $r=T_\odot\,m_\mathrm{B}$. The constant $T_\odot \equiv (\mathcal{GM})_\odot^\mathrm{N}/c^3$, where $c$ is the speed of light in vacuum and $(\mathcal{GM})_\odot^\mathrm{N} \equiv 1.327\,124\,4\times10^{26}\,  \mathrm{cm^3\,s^{-2}}$ is the nominal solar mass parameter defined by the IAU 2015 Resolution B3 \citep{Prsa2016}. Through out the paper, all masses expressed in solar mass $\mathrm{M_\odot}$ are referred to the nominal solar mass by taking the ratio $Gm_i/(\mathcal{GM})_\odot^\mathrm{N}\,(i=A,B)$, where $G$ is the gravitational constant.

The leading-order expression Eqs.~\eqref{eq:s_lo} and \eqref{eq:shapiro} were obtained by integrating along a straight line (in harmonic coordinates) and assuming a static mass distribution when the pulsar's signal propagates through the system \citep{BT1976}. 
In reality, the pulsar's signal propagates along a curved path due to the deflection in the gravitational field of the companion and leads to a \emph{lensing correction} to the Shapiro delay. This actually results in a reduced propagation time as a consequence of Fermat's principle \citep{Perlick2004}. 
The effect of lensing is not yet observable in any pulsar systems, but for completeness, one can extend Eq.~\eqref{eq:shapiro} by an adapted version of the approximation in \citet[Eq.~73]{KZ2010}: $\Lambda_u \rightarrow \Lambda_u + \delta \Lambda_u^\mathrm{len}$ with
\begin{align}
    \delta\Lambda_u^\mathrm{len} = 2rc/a_\mathrm{R} \,, \label{eq:len}
\end{align} 
where the semi-major axis of the relative orbit $a_\mathrm{R} = (x+x_\mathrm{B})/s$, with $x$ and $x_\mathrm{B}$\footnote{$x_\mathrm{B}$ had been observed in \citet{Kramer+2006}.} being the projected semi-major axes of pulsar A and pulsar B, respectively. For the Double Pulsar, one needs to account for the fact that the companion star moves while the pulsar's signal propagates across the system. This effect is known as \emph{retardation effect} or 1.5PN correction to the Shapiro delay \citep{KS1999,RL2006_lensing}. To sufficient approximation, the signal propagation delay can be extended to 
\begin{equation}
    \Delta_\mathrm{S} = -2r \ln{(\Lambda_u + \delta \Lambda_u^\mathrm{len} + \delta \Lambda_u^\mathrm{ret})} \,,
\end{equation}
where the retardation correction $\delta\Lambda_u^\mathrm{ret}$ can be taken directly from \citet[Eq.~130]{KS1999} as 
\begin{align}
    \delta\Lambda_u^\mathrm{ret} = &\frac{2\pi}{s} \frac{x}{P_\mathrm{b}}\frac{m_\mathrm{A}}{m_\mathrm{B}} e_\mathrm{T} \sin{u} - \frac{2\pi \,s}{(1-e_\mathrm{T}^2)^{1/2}} \frac{x}{P_\mathrm{b}} \frac{m_\mathrm{A}}{m_\mathrm{B}} \nonumber\\
    &\left[\sin{\omega}\, (\cos{u} - e_\mathrm{T}) + (1-e_\mathrm{T}^2)^{1/2} \cos{\omega} \sin{u} \right] \nonumber\\
    &\left[e_\mathrm{T} \cos{\omega} + \frac{(\cos{u -e_\mathrm{T}}) \cos{\omega} - (1-e_\mathrm{T}^2)^{1/2} \sin{\omega} \sin{u}}{1-e_\mathrm{T} \cos{u}} \right] \,.
    \label{eq:ret}
\end{align}
The quantity $P_\mathrm{b}$ denotes the orbital period, and $m_\mathrm{A}$ denotes the mass of pulsar A. Note, in the Double Pulsar, the mass ratio $m_\mathrm{A}/m_\mathrm{B}$ can be obtained in a theory-independent way \citep{Kramer+2006,Damour2007}. Hence, apart from the Shapiro shape parameter $s$, Eq.~\eqref{eq:ret} contains only Keplerian parameters.
\begin{figure}[t]
    \centering
    \includegraphics[width=\columnwidth]{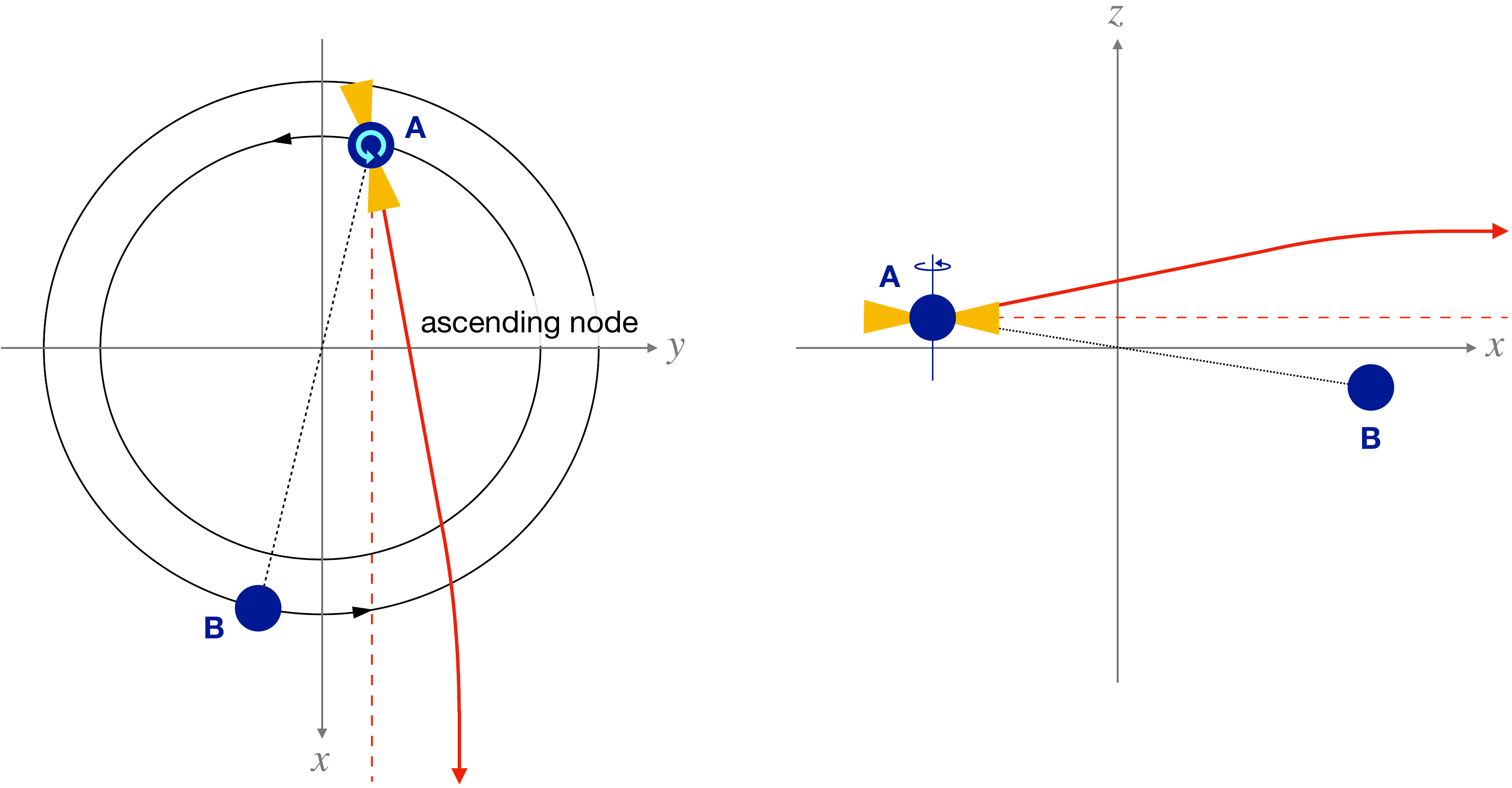}
    \caption{Simplified illustration of effects related to the deflection of A's radio signals (solid red) in the gravitational field of B (top down and side perspective). The observer is located at large distance along the x-axis. Apart from modifications in the propagation time due to a curved path in the gravitational field of B (lensing), one has a longitudinal deflection delay ($\delta_\mathrm{A}^\mathrm{londef}$) due to the fact that the pulsar has to rotate by more than 360$^\circ$ between two pulses while approaching the conjunction. After conjunction, it is less than 360$^\circ$, which makes pulsar signals arrive earlier at the observer. In addition, there is a latitudinal effect, due to a latitudinal shift in the emission direction towards the observer. This can lead to changes in the pulse profile since the line of sight cuts a different part of the emission region, which can also be accompanied by changes in the pulse arrival times (more details in Sections \ref{sec:profile} and \ref{sec:lat_time}).}
    \label{fig:defldelays}
\end{figure}

Moreover, the classical aberration expression \citep{SB1976} assumes a flat spacetime for the propagation of the pulsar signals, which is no longer sufficient for describing the observations of the Double Pulsar, particularly near the superior conjunction of pulsar A. One needs to account for the gravitational deflection of the pulsar's signal caused by its companion \citep{DK1995,RL2006_rotation}, which adds a lensing correction to the classical aberration. For pulsar A the misalignment angle between its spin vector and the orbital angular momentum is very small \citep[$<3.2 \si\degree$,][]{Ferdman+2008,Ferdman2013}, which is in line with a low-kick birth event \citep[cf.][]{Piran+2004,Willems+2004, Willems+2006,Ingrid+2006,Tauris+2017}. Since the spin of A is practically parallel to the orbital angular momentum, the aberration delay can be simplified as 
\begin{align}
    \Delta_\mathrm{A} &= \mathcal{A}\, \Bigl(\sin{\psi} + e_\mathrm{T}\sin{\omega}\Bigr) + \delta_\mathrm{A}^\mathrm{\,londef} \label{eq:aber} \,.
\end{align}
The first term on the right-hand side of Eq.~\eqref{eq:aber} is the classical aberration delay, where $\psi=\omega+\theta$ is the longitude of pulsar with respect to the ascending node ($\theta$ is the true anomaly, which defines the angle between the direction of the pulsar and the periastron), and the aberration coefficient
\begin{align}
    \mathcal{A} = \frac{x}{\nu P_\mathrm{b} (1-e_\mathrm{T}^2)^{1/2} \sin^{2}{i}} \simeq 3.65\,\mu\text{s} \,.\label{eq:A0}
\end{align}
As $\mathcal{A}$ is practically not observable and can be absorbed by a shift in various timing parameters \citep[see discussions in][]{DD86,DT92}, we \emph{a priori} add the aberration coefficient $\mathcal{A}$ as a fixed parameter in our timing model with the value given in Eq.~\eqref{eq:A0}.

The second term  $\delta_\mathrm{A}^\mathrm{\,londef}$ in Eq.~\eqref{eq:aber} is the higher-order correction originating from the \emph{longitudinal deflection delay}, and can be written as \citep{DK1995}
\begin{align}
    \delta_\mathrm{A}^\mathrm{\,londef} &= \mathcal{D}\, \frac{\cos{(\psi+
    \delta\psi^\mathrm{ret})}}{\Lambda_u + \delta \Lambda_u^\mathrm{ret}}\,,\text{with}\,\, \mathcal{D} = \frac{1}{\pi \nu}\frac{r}{x + x_\mathrm{B}} \,.
    \label{eq:defl}
\end{align}
Like in the Shapiro delay, retardation correction is also accounted for here. As a sufficiently good approximation, the position of B when the signal reaches its minimum distance from B can be used \citep[retardation corrected position; cf.][]{KS1999,RL2006_lensing}. The angle $\delta\psi^\mathrm{ret}$ denotes the retardation related correction for the angle between the (coordinate) vector from B to A and the ascending node.

As already discussed in \citet{Kramer+2021DP}, the NLO contributions in the Shapiro and aberration delays can not be tested separately in the Double Pulsar due to the similarity of their effects on signal propagation. In addition, the lensing correction to the propagation delay (Eq.~\ref{eq:len}) is challenging to measure as it can be absorbed in the fit of Shapiro shape $s$ \citep[see Section~\ref{sec:lensing} and discussions in][]{Kramer+2021DP}. Therefore, to test the significance of the NLO contributions and to obtain an unbiased timing result, a common factor $q_\mathrm{NLO}$ is multiplied by these contributions and can be fitted for in our timing model:
\begin{align}
    \Lambda_u^\mathrm{ret} &= \Lambda_u^\mathrm{ret} \times q_\mathrm{NLO} \,,\label{eq:ret_nlo}\\
    \Lambda_u^\mathrm{len} &= \Lambda_u^\mathrm{len} \times q_\mathrm{NLO} \,,\label{eq:len_nlo}\\
    \delta_\mathrm{A}^\mathrm{\,londef} &= \delta_\mathrm{A}^\mathrm{\,londef} \times q_\mathrm{NLO} \,.\label{eq:def_nlo}
\end{align}
In GR, the scaling factor $q_\mathrm{NLO}=1$. Figure~\ref{fig:defldelays} illustrates the different effects related to signal deflection in the Double Pulsar system.

\section{Timing results}
\label{sec:timing}
For the timing analysis, we use the timing model in \textsc{tempo} known as DDS, which is a modification of the DD model \citep{DD86} that uses a different parameterisation of the Shapiro delay. In DDS, the Shapiro shape parameter $s$ is replaced by the logarithmic Shapiro shape parameter $z_s$ via 
\begin{equation}
    z_s\equiv -\ln{(1-s)},
    \label{eq:zs}
\end{equation}
which is more suitable when $s$ is very close to 1 \citep[see][]{Kramer+2006AnP, Kramer+2021DP}. 
The NLO contributions in the Shapiro and aberration delays are also implemented in the latest DDS model, which can be measured through a common factor $q_\mathrm{NLO}$. Because the analytic inversion of the timing model developed in \citet{DD86} is no longer sufficient for the Double Pulsar, primarily due to the NLO contributions, a numerical inversion of the timing model was also implemented in the latest DDS model in \textsc{tempo} \citep[see][]{Kramer+2021DP}.

\subsection{Timing parameters}
\label{sec:timing_par}
\begin{figure}[t]
    \centering
    \includegraphics[width=\columnwidth]{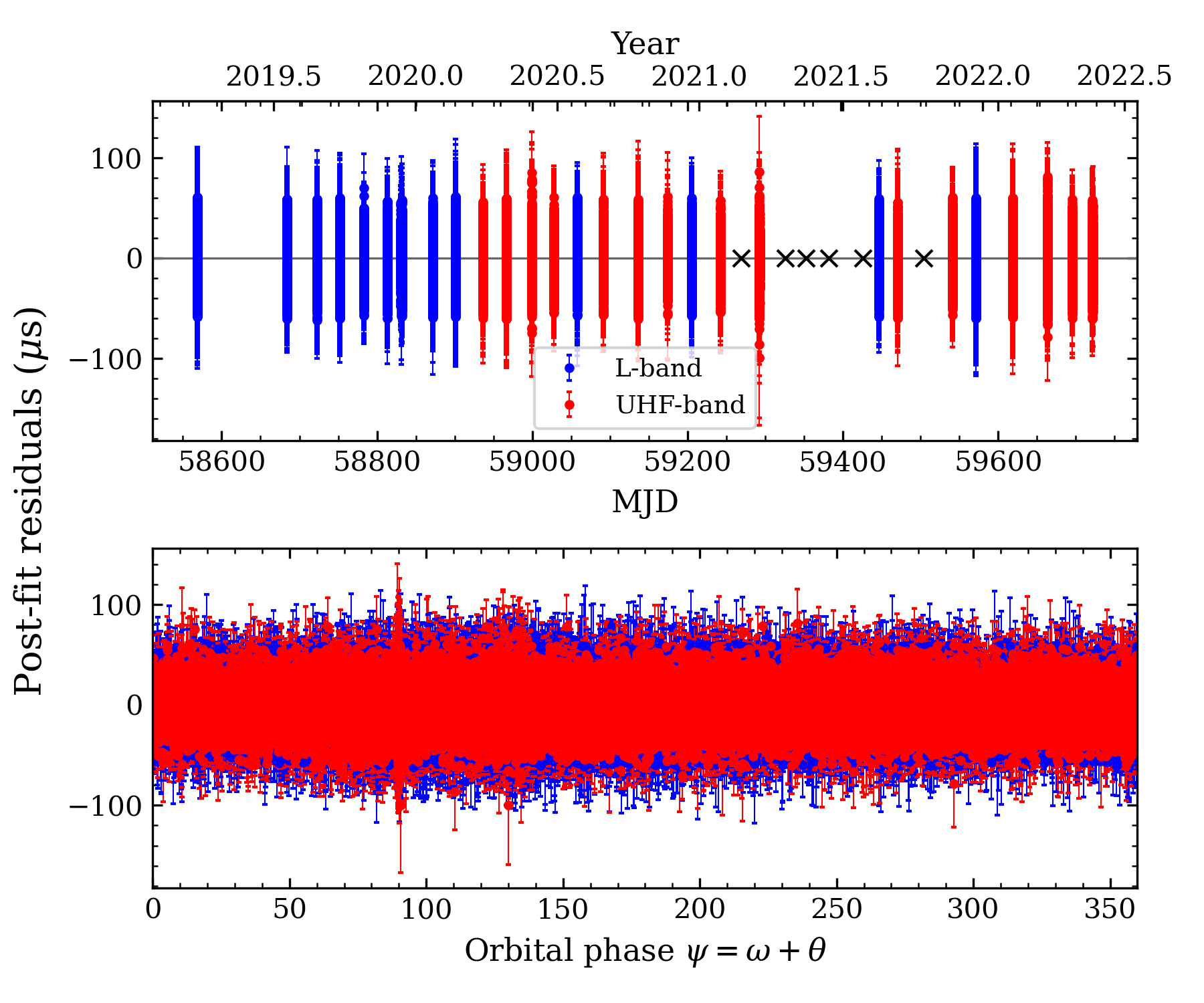}
    \caption{Post-fit residuals of PSR~J0737$-$3039A using the DDS binary model as a function of time (top panel) and orbital phase of pulsar A with respect to the ascending node $\psi$ (bottom panel). The MeerKAT L-band data are plotted in blue, whereas the UHF-band data are in red. The epochs of the excluded 6 observations are marked as black crosses.}
    \label{fig:res}
\end{figure} 

In our analysis, the full MeerKAT data set shows a large deviation in the Shapiro range parameter $r$ compared to the 16-yr result \citep{Kramer+2021DP}. We perform a drop-out analysis by removing each observing epoch and fitting the parameters. We find that $r$ is dependent on specific observing epochs, where 6 epochs affect $r$ by a significant amount while the rest epochs don't. These 6 epochs are marked as black crosses in Fig.~\ref{fig:res}.
After excluding all these 6 epochs, $r$ is consistent with the 16-yr result and the mass measurement in GR \citep{Kramer+2021DP}. Even though a number of tests and simulations have been made, we are still unclear about the cause of this problem. Possible reasons could be systematic errors in the observations or folding techniques. 
Note that all data were folded with \textsc{tempo2} phase predictor during observations, which has shown outliers in this pulsar and has been doubly confirmed by our simulations. These outliers disappear after reinstalling a \textsc{tempo} \texttt{polyco} ephemeris in data processing (see Section~\ref{sec:data_reduce}), but we cannot rule out underlying problems due to folding technique. The results shown here are based on data processed with \texttt{polyco} scheme\footnote{With the same set of observations, data processed with \textsc{tempo} \texttt{polyco} and \textsc{tempo2} \texttt{predictor} show a noticeable difference ($\sim3\sigma$) in the Shapiro parameters, where the result with \texttt{polyco} is closer to the 16-yr results and shows a smaller $\chi^2$.}. 
Anyway, this issue should not affect the measurement of NLO signal propagation effects, which is the main focus of this paper. Therefore, we leave this question to future studies. In the following analysis, we use a trimmed data set which excludes these 6 epochs.

Table~\ref{tab:dds} and Fig.~\ref{fig:res} present the results obtained from fitting the \textsc{tempo} DDS model to the trimmed MeerKAT data set.
We fix the proper motion ($\mu_\alpha, \mu_\delta$) and parallax $\pi_\mathrm{x}$ to the more precise values determined from the 16-yr timing and VLBI measurement \citep[see][]{Kramer+2021DP}. 
As the time span of our data is not sufficient to obtain a reliable measurement of the orbital period derivative $\dot{P}_\mathrm{b}$ and the orientation of the orbit ($\omega_0 \approx 0^\circ$) is not at a favourable position for a precise measurement of the Einstein delay amplitude $\gamma_\mathrm{E}$, we choose to fix these parameters to the more precise measurements from 16-yr data \citep{Kramer+2021DP}. Fixing the above parameters has no impact on the measurements of signal propagation effects and masses.
The two PK parameters that describe the relativistic deformation of the orbit, $\delta_r$ and $\delta_\theta$ \citep{DD86}, are also held fixed at the GR value in our analysis, as $\delta_r$ cannot be measured \citep[see][]{Kramer+2021DP} and $\delta_\theta$ is not yet measurable with the current MeerKAT data.

{\renewcommand{\arraystretch}{1.2}
\begin{table}[t]
\caption{Timing parameters for PSR~J0737$-$3039A using \textsc{tempo} DDS binary model. Numbers in parentheses are $1\sigma$ uncertainties referred to the last digits, obtained from the standard deviation of 1000 MC runs or maximum error from \textsc{tempo}, whichever is larger. The overall reduced $\chi^2$ is 0.99.}
\centering
\begin{threeparttable}
\begin{tabular}{lc}
\hline \hline
\textbf{Parameter} &  \textbf{Value} \\ \hline
Solar System ephemeris & DE436 \\
Terrestrial time standard & UTC(NIST) \\
Timescale & TDB \\
Position epoch (MJD) & 55045.0 \\
\vspace{5pt}
Timing epoch, $t_0$ & 55700.0 \\
\emph{Astrometric parameters} \\
Right ascension (R.A.), $\alpha$ (J2000) & 07:37:51.248\,121(26) \\
Declination (Dec.), $\delta$ (J2000) & $-$30:39:40.705\,36(42) \\
Proper motion R.A., $\mu_\alpha$ (mas yr$^{-1}$) & $-$2.567(30)$^*$ \\
Proper motion Dec., $\mu_\delta$ (mas yr$^{-1}$) & 2.082(38)$^*$\\
\vspace{5pt}
Parallax, $\pi_\mathrm{x}$ (mas) & 1.36($+$0.12,$-$0.10)$^*$\\
\emph{Spin parameters} \\
Rotational frequency (freq.), $\nu$ (Hz) & 44.054\,068\,642\,001(56) \\
First freq. derivative, $\dot{\nu}\, (\mathrm{Hz\,s^{-1}})$ & $-3.415\,92(37)\times 10^{-15}$\\
\vspace{5pt}
Second freq. derivative, $\ddot{\nu}\, (\mathrm{Hz\,s^{-2}})$ & $-9.5(12) \times 10^{-27}$\\
\emph{Binary parameters} \\
Orbital period, $P_{\text{b}}$ (days)  & 0.102\,251\,559\,297\,2(29)\\
Projected semi-major axis, $x$ (s) & 1.415\,028\,299(88)\\
Eccentricity, $e_\mathrm{T}$ & 0.087\,777\,036(48)\\
Epoch of periastron, $T_{0}$ (MJD) & 55700.233\,017\,54(10)\\
Longitude of periastron, $\omega$ (deg) & 204.753\,72(36)\\
Periastron advance, $\dot{\omega}$ (deg yr$^{-1}$) & 16.899\,321(37)\\
Orbital period derivative ($10^{-12}$), $\dot{P}_\mathrm{b}$ & $-1.247\,920(78) ^*$\\
Einstein delay amplitude, $\gamma_\mathrm{E}$ (ms)& 0.384\,045(94)$^*$\\
Logarithmic Shapiro shape, $z_s$ & 9.669(77)\\
Range of Shapiro delay, $r$ ($\mu$s) & 6.163(16) \\
\vspace{5pt}
NLO factor for signal prop., $q_\mathrm{NLO}$ & 0.999(79) \\
\emph{Derived parameters} \\
$s\equiv\sin{i}=1-e^{-z_s}$ & 0.999\,936\,9(+46/ -51)\\
Orbital inclination, $i$ (deg) & 89.36(3) or 90.64(3)\\
Mass of pulsar A, $m_\mathrm{A}\mathrm{(M_\odot)}$ & 1.338\,186(10) \\
Mass of pulsar B, $m_\mathrm{B}\,\mathrm{(M_\odot)}$ & 1.248\,866(7) \\
Total mass, $M\,\mathrm{(M_\odot)}$ & 2.587\,052(11) \\
\hline
\end{tabular}
\begin{tablenotes}
$^*$ Values adopted from \citet{Kramer+2021DP}.
\end{tablenotes}
\end{threeparttable}
\label{tab:dds}
\vspace{-5pt}
\end{table}
}

The values shown in Table~\ref{tab:dds} are the result of 1000 Monte-Carlo (MC) runs, where in each run, a random realisation of proper motion, parallax, DM, $\gamma_\mathrm{E}$, and $\dot{P}_\mathrm{b}$ is selected. The DM value is selected according to the DM measurements and uncertainties shown in Fig.~\ref{fig:dm}. We use the aforementioned modified version of \textsc{tempo} to correct DM for each TOA and fit for all other timing parameters in each run.
The numbers shown in Table~\ref{tab:dds} are the mean values of the distribution of each parameter after 1000 MC runs, whereas the uncertainties is taken from the larger one among the standard deviation of the distributions and the maximum error from \textsc{tempo} in all MC runs.

In order to allow direct comparisons with previous publications, parameters shown in Table~\ref{tab:dds} are measured with respect to the same epochs and the same terrestrial time standard UTC(NIST)\footnote{ \url{https://www.nist.gov/pml/time-and-frequency-division/time-realization/utcnist-time-scale-0}.} within the timescale ``Barycentric Dynamical Time (TDB)'' as implemented in \textsc{tempo}. Even though TDB runs at a slower rate than the ``Barycentric Coordinate Time (TCB)'', which was recommended by IAU 2006 Recolution B3\footnote{\url{https://www.iau.org/static/resolutions/IAU2006_Resol3.pdf}} \citep[see also][]{IAU2003}, this choice does not have any impact on the results presented in this paper \citep[see discussions in][]{Kramer+2021DP}.
To transfer the TOAs from topocentric to the Solar System Barycentre (SSB), the DE436 Solar System ephemeris published by the Jet Propulsion Laboratory\footnote{\url{https://ssd.jpl.nasa.gov/planets/eph_export.html}} is used. 

All binary parameters in Table~\ref{tab:dds} are consistent with the 16-yr data, except for $x$ being  different by $\sim 3\sigma$. This is because $x$ is highly correlated with $\gamma_\mathrm{E}$ which is kept fixed in our fit. This should be improved in the future once we have enough MeerKAT data to fit for $x$ and $\gamma_\mathrm{E}$ simultaneously.
In our fit, the root mean square (RMS) is very close to the mean TOA uncertainty, and the reduced $\chi^2$ of individual observation is close to 1, suggesting that our result is not affected by jitter noise. We also perform simulations and single-pulse analysis following the methods in \citet{Parthasarathy+2021} and find little evidence of jitter noise.

The RMS of the MeerKAT data shown in Table~\ref{tab:compare} is more than two times better than that of the Green Bank Telescope \citep[see][]{Kramer+2021Relbin}. Thanks to this much improved precision, the measurements of the Shapiro parameters improve quickly. Compared to \citet{Kramer+2021DP}, the shape parameter $s$ improves by a factor of 2 and the range parameter $r$ improves by a factor of 1.3 (see Table~\ref{tab:dds}).

{\renewcommand{\arraystretch}{1.2}
\begin{table}
\caption{Mass measurements with a new modified DDGR model which accounts for NLO contributions in the orbital motion and signal propagation in this system. The MOI has been chosen to be $I_\mathrm{A}=1.28\times 10^{45} \mathrm{g\,cm^{2}}$ in the fit.}
\centering
\begin{tabular}{lc}
\hline \hline
\textbf{Parameter} &  \textbf{Value} \\ \hline
Mass of pulsar A, $m_\mathrm{A}\,\mathrm{(M_\odot)}$ & 1.338\,186(10)\\
Mass of pulsar B, $m_\mathrm{B}\,\mathrm{(M_\odot)}$ & 
1.248\,886(5)\\
Total mass, $M\,\mathrm{(M_\odot)}$ & 
2.587\,050(8)\\
\hline
\end{tabular}
\label{tab:ddgr}
\end{table}
}

\subsection{Mass measurements}
The standard approach for measuring the masses of a binary pulsar system is using two PK parameters. Assuming GR, one can calculate the two \emph{a priori} unknown masses based on the measurements of Keplerian parameters. For the Double Pulsar, the two most precisely measured PK parameters are periastron advance $\dot{\omega}$ and the Shapiro shape parameter $s$. 

For the advance of periastron, in addition to the 1PN contribution, we also need to account for higher-order corrections due to 2PN effects and Lense-Thirring (LT) precession caused by spin-orbit coupling of pulsar A, as they are much larger than the measurement error of $\dot{\omega}$ \cite[see][]{Hu2020,Kramer+2021DP}.
For the analysis of this paper, the total intrinsic contribution to the periastron advance can be expressed, with sufficient precision, as \citep{DS88}
\begin{align}
    \dot{\omega} = \dot{\omega}^\mathrm{1PN} + \dot{\omega}^\mathrm{2PN} +\dot{\omega}^\mathrm{LT,A} \,.
    \label{eq:omdot}
\end{align}
The first and second post-Newtonian (PN) terms $\dot{\omega}^\mathrm{1PN}$ and $\dot{\omega}^\mathrm{2PN}$ are functions of masses and observed Keplerian parameters. While the situation is more complicated for the LT contribution $\dot{\omega}^\mathrm{LT,A}$, as it requires the knowledge of the moment of inertia (MOI) of pulsar A, which is still not very constrained because of our limited knowledge of the equation of state (EOS) of neutron stars. As discussed in \citet{Hu2020} and \citet{Kramer+2021DP}, one could measure the masses and the MOI simultaneously by introducing a third PK parameter $\dot{P}_\mathrm{b}$ into the $\dot{\omega}-s$ test. Such a test have already been made using the 16-yr data with an upper limit obtained: $I_\mathrm{A}<3.0\times 10^{45}\ \mathrm{g\,cm^{2}}$ with 90\% confidence \citep{Kramer+2021DP}. This measurement is expected to improve with the combination of the 16-yr data with MeerKAT data in a forthcoming paper, and should improve considerably over the next years with more data taken. This promises an important complementary constraint on the EOS \citep{Hu2020}.
For the calculations here, we take the value from \citet[Eq.~35]{Kramer+2021DP} which uses the constraints on the EOS from \citet{Dietrich+2020}: 
\begin{align}
    \dot{\omega}^\mathrm{LT,A} = -4.83(+29,-35) \times 10^{-4} \;\text{deg yr}^{-1} \,.
    \label{eq:LT}
\end{align}

The Shapiro shape parameter $s$ is the sine of the orbital inclination $i$. 
In Newtonian gravity, the orbital inclination is linked to the projected semi-major axis $x$ via the binary mass function \citep[e.g.][]{LK2004}:
\begin{align}
    \sin{i} = \frac{(n_\mathrm{b}M)^{2/3} x }{T_\odot^{1/3} m_\mathrm{B}}\,,
    \label{eq:sini}
\end{align}
where $x$ and the orbital frequency $n_\mathrm{b}\equiv 2\pi/P_\mathrm{b}$ are both measured Keplerian parameters. Eq.~\eqref{eq:sini} gets modified by a 1PN term in the 1PN approximation for Kepler's third law (see Eq.~3.7 in \citeauthor{DD85} \citeyear{DD85} and Eq.~3.9 in \citeauthor{DT92} \citeyear{DT92}):
\begin{align}
    \sin{i} = \frac{(n_\mathrm{b}M)^{2/3} x }{T_\odot^{1/3} m_\mathrm{B}} \left[1+ \left(3-\frac{m_\mathrm{A}m_\mathrm{B}}{3 M^2}\right) \Big(T_\odot M n_\mathrm{b} \Big)^{2/3} \right]\,.
    \label{eq:sini1PN}
\end{align}
Taking the measurements of $P_\mathrm{b}$, $x$, and masses from Table~\ref{tab:dds}, one can calculate that the 1PN correction is approximately $1.27\times 10^{-5}$. This correction was considered for the first time in pulsar analysis by \citet{Kramer+2021DP}, where the significance was about $1.3\sigma$. Now with MeerKAT data, this 1PN correction is $2.5\sigma$ significant and can not be ignored in the analysis. We hereby use the full 1PN mass function Eq.~\eqref{eq:sini1PN} to measure the masses.

Combining the PK parameters $\dot{\omega}$ and $s$, one obtains the (Doppler-shifted) masses, which are listed in Table~\ref{tab:dds}. These measurements are fully consistent with those obtained with 16-yr data \citep{Kramer+2021DP}, and the precision of $m_\mathrm{A}$ and $m_\mathrm{B}$ are also better.

Alternatively, one can fit for masses using the timing model known as DDGR \citep{TW1989}, which is based on the DD model where the PK parameters are explicitly calculated from the masses and the Keplerian parameters assuming GR. Beside the Keplerian parameters, it fits explicitly for the total mass $M$ and the companion mass $m_\mathrm{B}$. 
To make the measurements, we modify the DDGR model so that it incorporates all NLO contributions that need to be accounted for in this system, including NLO signal propagation effects, LT contribution $\dot{\omega}^\mathrm{LT,A}$, NLO gravitational wave damping and mass loss contribution to $\dot{P}_\mathrm{b}$ \citep[see][]{Hu2020, Kramer+2021DP}. An MOI needs to be provided to the model for the calculation of $\dot{\omega}^\mathrm{LT,A}$ and the mass loss contribution to $\dot{P}_\mathrm{b}$. For periastron advance $\dot{\omega}$, the uncertainty from the MOI is still smaller than that from MeerKAT observations (see Eq.~\ref{eq:LT} and Table~\ref{tab:dds}). Therefore, based on the EOS constraint from \citet{Dietrich+2020}, we fix the MOI to $I_\mathrm{A}=1.28\times 10^{45} \mathrm{g\,cm^{2}}$ in our fit.
Table~\ref{tab:ddgr} shows the mass measurements obtained using the DDGR model. The results are fully consistent with the measurements derived from the DDS model, with smaller uncertainties in $m_\mathrm{B}$ and $M$.

Following \citet{Kramer+2021DP}, one could test the agreement of $r$ in GR by comparing $m_\mathrm{B}^{(r)} = r/T_\odot=1.2512(33)\,\mathrm{M_\odot}$ (cf. Table~\ref{tab:dds}) with the companion mass determined here, which gives
\begin{equation}
    r^\mathrm{obs}/r^\mathrm{GR} = 1.0019(26) \,.
\end{equation}
This leads to a $5.3\times 10^{-3}$ ($95\%$ confidence) test of GR. 
\begin{figure}[t]
    \centering
    \includegraphics[width=0.95\columnwidth]{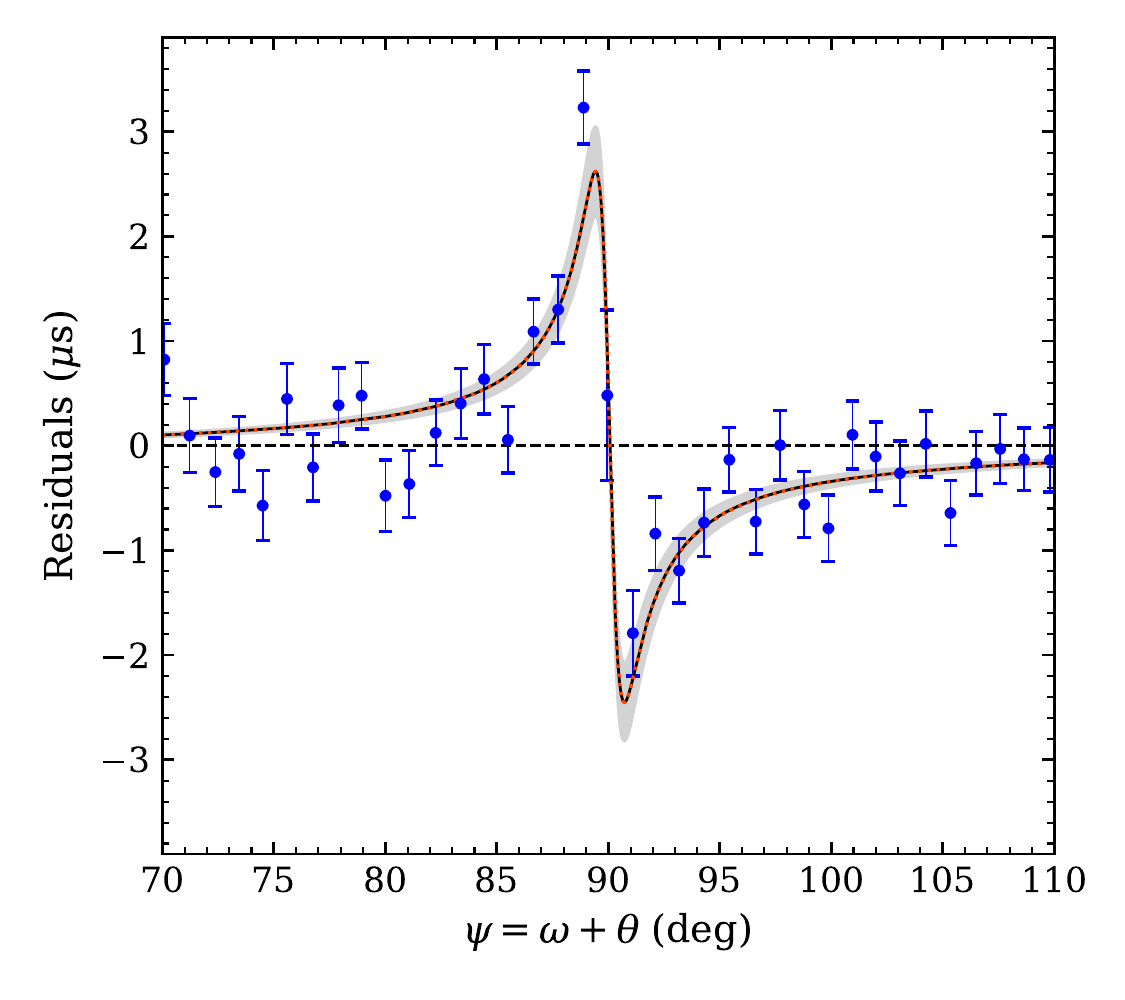}
    \vspace{-5pt}
    \caption{Aggregated residuals (blue) due to NLO contributions in Shapiro and aberration delay, shown in the orbital phase $\psi$. 
    Residuals are re-scaled by $(1+e_\mathrm{T}\cos{\theta})^{-1}$ to account for secular variations in amplitude due to the precession of periastron. The black curve indicates the fitted $q_\mathrm{NLO}$ (see Table~\ref{tab:dds}) with the $2\sigma$ range shown in grey shadow, which agrees very well with the theoretical prediction indicated by the red dotted line.}
    \label{fig:shaphof}
\end{figure}

\section{Studying NLO signal propagation effects}
\label{sec:test_NLO}

Because the Double Pulsar system is nearly edge-on to our line-of-sight (LOS, see $i$ in Table~\ref{tab:dds}), it is ideal for measuring signal propagation effects caused by the gravitational field of the companion near superior conjunction. The leading-order expression Eq.~\eqref{eq:s_lo} is no longer sufficient to describe the signal propagation in the Double Pulsar. Such a model would result in significant residuals near superior conjunction when aggregating residuals in the orbital phase, as shown in Fig.~\ref{fig:shaphof}. These residuals agree very well with the expected NLO contributions discussed in Section~\ref{sec:NLO}, which is shown by the red curve. The significance of the NLO corrections can be tested by scaling these corrections collectively by a common factor $q_\mathrm{NLO}$ (cf. Eqs.~\ref{eq:ret_nlo}-\ref{eq:def_nlo}; $q_\mathrm{NLO}$=1 in GR) and fit for it. We find, with the trimmed data set,
\begin{align}
    q_\mathrm{NLO} = 0.999(79) \,,
\end{align}
which has surpassed the 16-yr result by 1.65 times with only $\sim$3\,yr of data thanks to the much improved precision offered by MeerKAT. 

Following the definition of $q_\mathrm{NLO}$ in Section~\ref{sec:NLO}, a fit for this parameter involves two aspects of gravity: 1.5PN correction of the Shapiro delay due to the movement of the companion $\delta \Lambda_u^\mathrm{ret}$, and corrections related to the signal deflection in the gravitational field of the companion $\delta \Lambda_u^\mathrm{len}$ and $\delta_\mathrm{A}^\mathrm{\,londef}$.
Even though these contributions can not be tested individually in a simultaneous fit because of their similarity, one can still test one at a time while keeping the other one fixed \citep[cf.][]{Kramer+2021DP}. We find
\begin{align}
    q_\mathrm{NLO}\text{[deflection]} &= 1.00(15) \,,\label{eq:q_def}\\
    q_\mathrm{NLO}\text{[retardation]} &= 1.00(17) \,.\label{eq:q_ret}
\end{align}

\subsection{Searching for profile variation at eclipse}
\label{sec:profile}

\begin{figure}[t]
    \centering
    \includegraphics[width=\columnwidth]{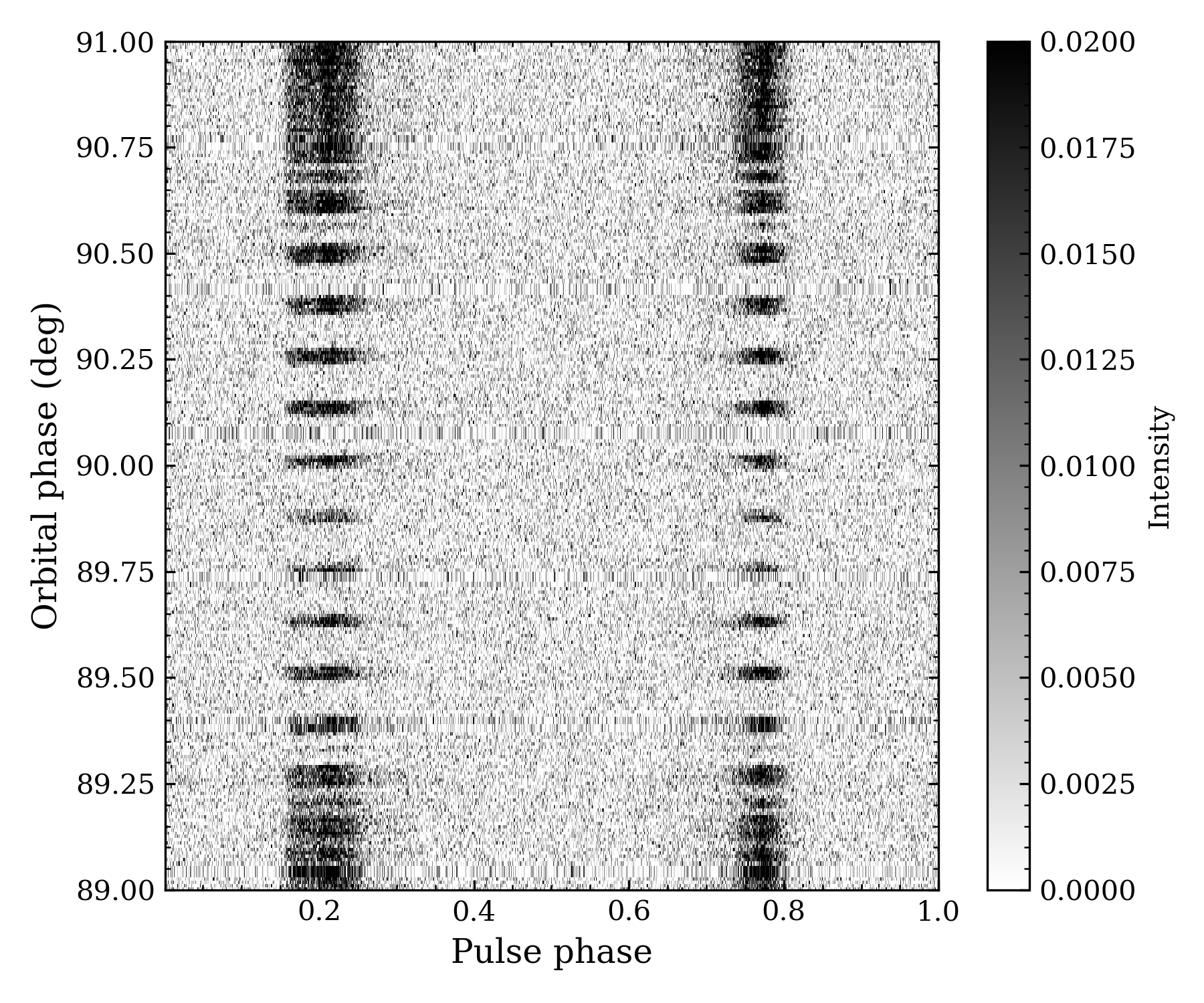}
    \vspace{-10pt}
    \caption{Example of one eclipse observation at UHF band, plotted in intensity against orbital phase $\psi$ and pulse phase. The intensity modulation occurs when pulsar A is eclipsed by the magnetosphere of pulsar B. Each integration is a sum of 8 pulses. When plotting, the discontinuities between recordings are patched with the previous sub-integration.}
    \label{fig:ecl}
\end{figure}
\begin{figure}[t]
    \centering
    \includegraphics[width=\columnwidth]{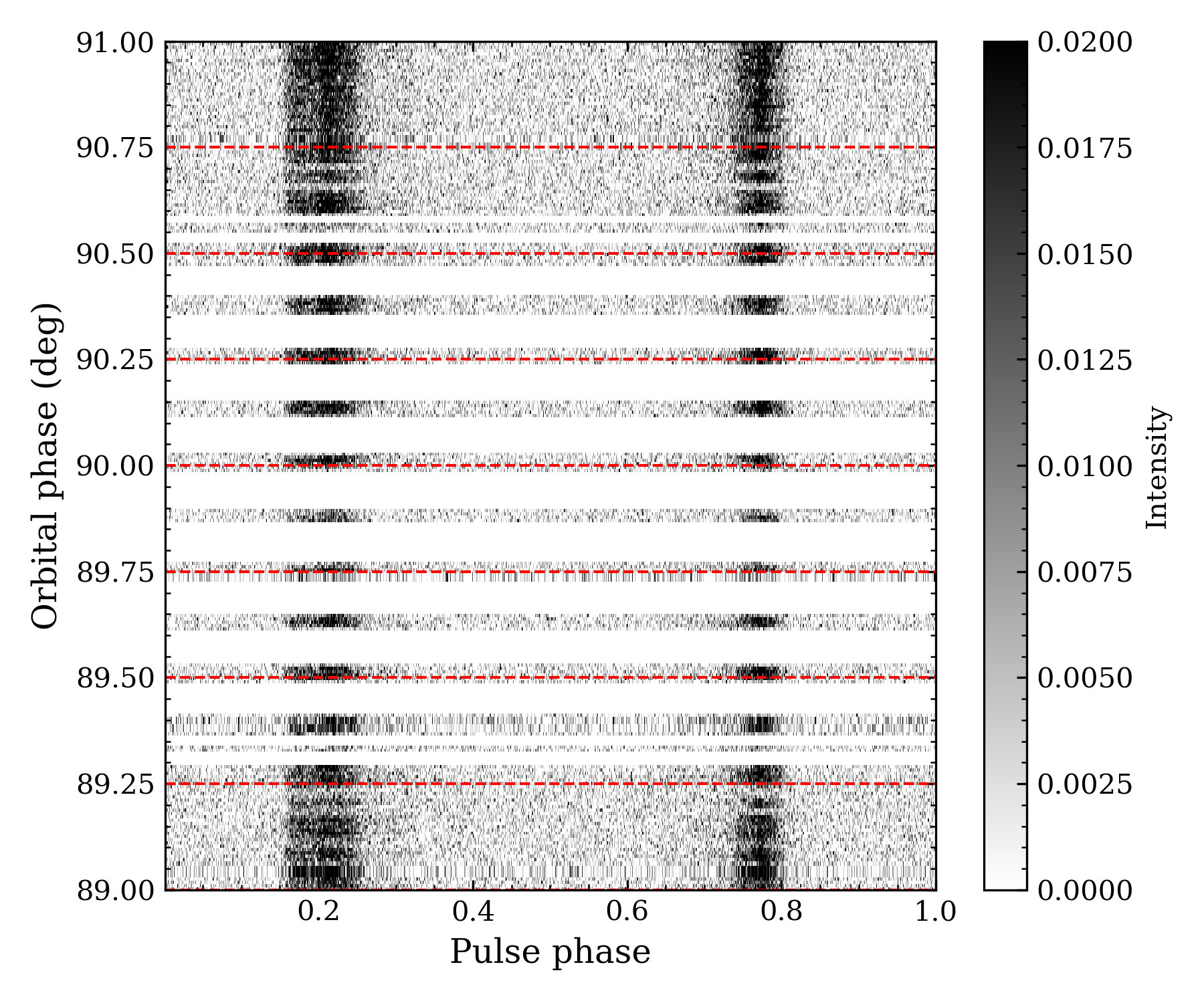}
    \vspace{-10pt}
    \caption{Same as Fig.~\ref{fig:ecl} but masking out the regions without pulses (blocked by the magnetosphere of pulse B). The red dashed lines indicate the orbital phase bins used in Figs.~\ref{fig:ecl_L} and \ref{fig:ecl_U}.}
    \label{fig:ecl_mask}
\end{figure}

The lensing correction to the aberration delay may not only lead to a shift in time in the longitudinal aspect but can also result in a change of the colatitude of the emission direction towards Earth, i.e., the \emph{latitudinal deflection delay}. This would cause profile variations as the LOS cuts a different region of the pulsar beam \citep{RL2006_lensing, RL2006_rotation}.
An illustration of the latitudinal deflection effect is shown on the right of Fig.~\ref{fig:defldelays}. Various analyses have confirmed that pulsar A is an orthogonal rotator \citep{Guillemot+2013, Ferdman2013, Kramer+2021Relbin}, that the main pulse and the interpulse come from opposite magnetic poles. 
Therefore, we do not expect shifts of pulse components in phase as discussed in \citet{RL2006_rotation} based on the (incorrect) assumption of an aligned rotator suggested by \citet{Demorest+2004}. 

The profile variation is expected to be maximum at the superior conjunction and symmetric around $\psi=90\si\degree$ (retardation corrected). This study requires high time resolution, for which we use the search mode data.
We select the data that are near the eclipses and fold them into single pulses using \textsc{tempo} \texttt{polyco} (with TSPAN=1min). Data are then combined, cleaned, and polarisation calibrated before integrated into total intensity and averaged in frequency.
As the single pulses are still very weak, we average over every 8 pulses to increase the single-to-noise (S/N). An example of eclipse data is shown in Fig.~\ref{fig:ecl}.

In order to get a high S/N profile, we first mask the regions where pulsar A's emission is blocked by pulsar B (see Fig.~\ref{fig:ecl_mask}), and split the data in orbital phase with a step of $\Delta\psi=0.25\si\degree$ for $89\si\degree <\psi < 91\si\degree$. 
Then for each phase interval, we integrate pulses from all observations of a given band (L or UHF) together to increase the S/N. The resulting profiles are shown in Fig.~\ref{fig:ecl_L} for the L-band data and in Fig.~\ref{fig:ecl_U} for the UHF-band data, which are summed from 25 and 37 eclipses respectively. The difference between the added profiles at the eclipse and a 2-hr integrated profile at the non-eclipse part of the orbit is insignificant. The subtle residual structures in these figures can result from interstellar medium effects (DM variation and scintillation) based on our simulations. 
Therefore, we conclude that the current data are not (yet) sensitive to profile variations caused by the latitudinal aberration delay, or are not significant in the region that is seen by our LOS. 
These profiles will be used in a subsequent study on the geometry of the system.

\afterpage{%
\begin{figure*}[t]
    \centering
    \includegraphics[width=0.9\textwidth]{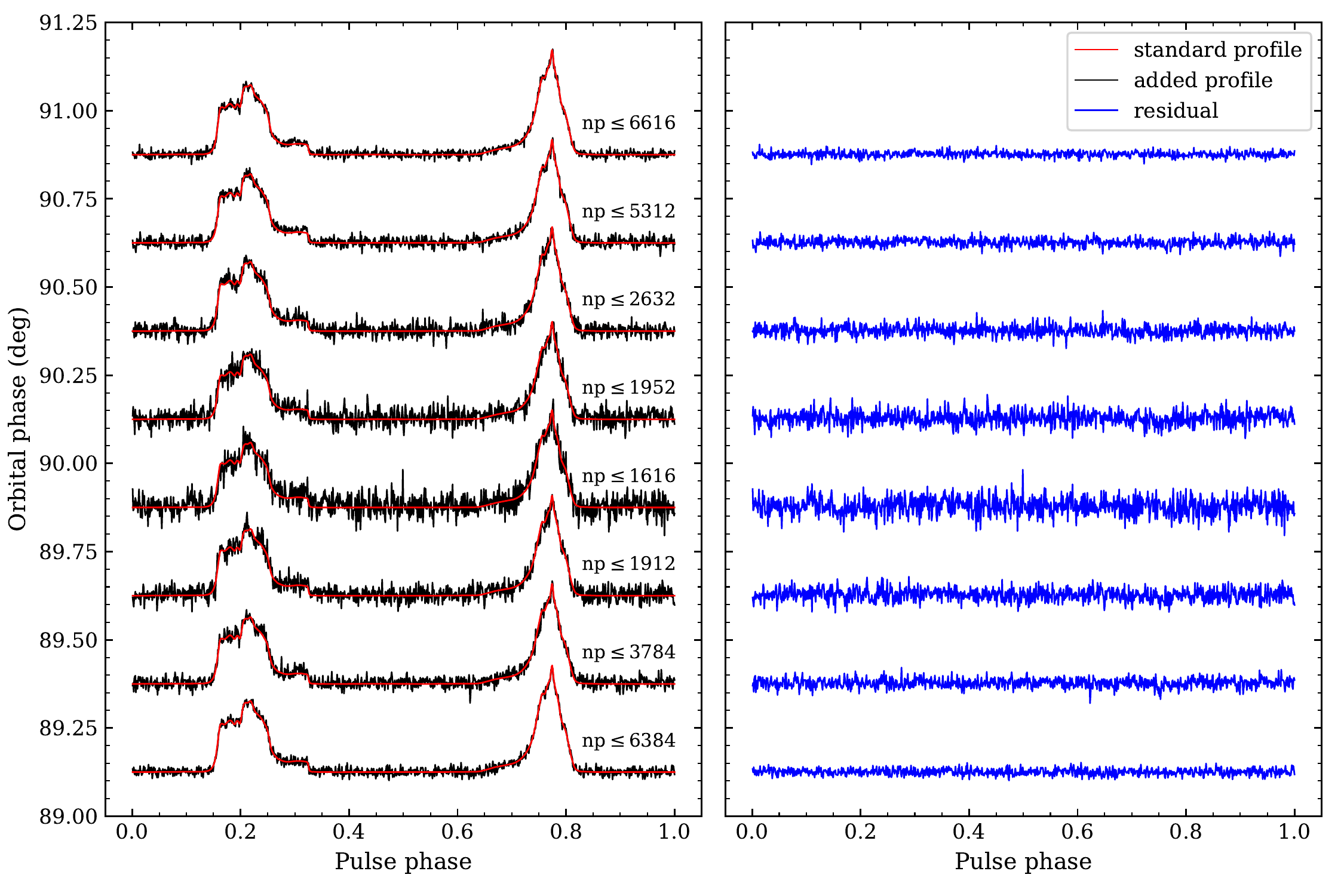}
    \caption{Left: The black lines indicate the integrated profile of L-band data over various orbital phase intervals summed from 25 eclipses. The red curves indicate the reference profile integrated over a 2-hr observation excluding the eclipse part. The baseline of each profile is placed at the central orbital phase of the interval, and the number on the right side of the profile (np) indicates the upper estimate of the number of pulses in that interval. Right: Residuals of the added profile compared to the reference profile.}
    \label{fig:ecl_L}
\end{figure*}
\begin{figure*}[t]
    \centering
    \includegraphics[width=0.9\textwidth]{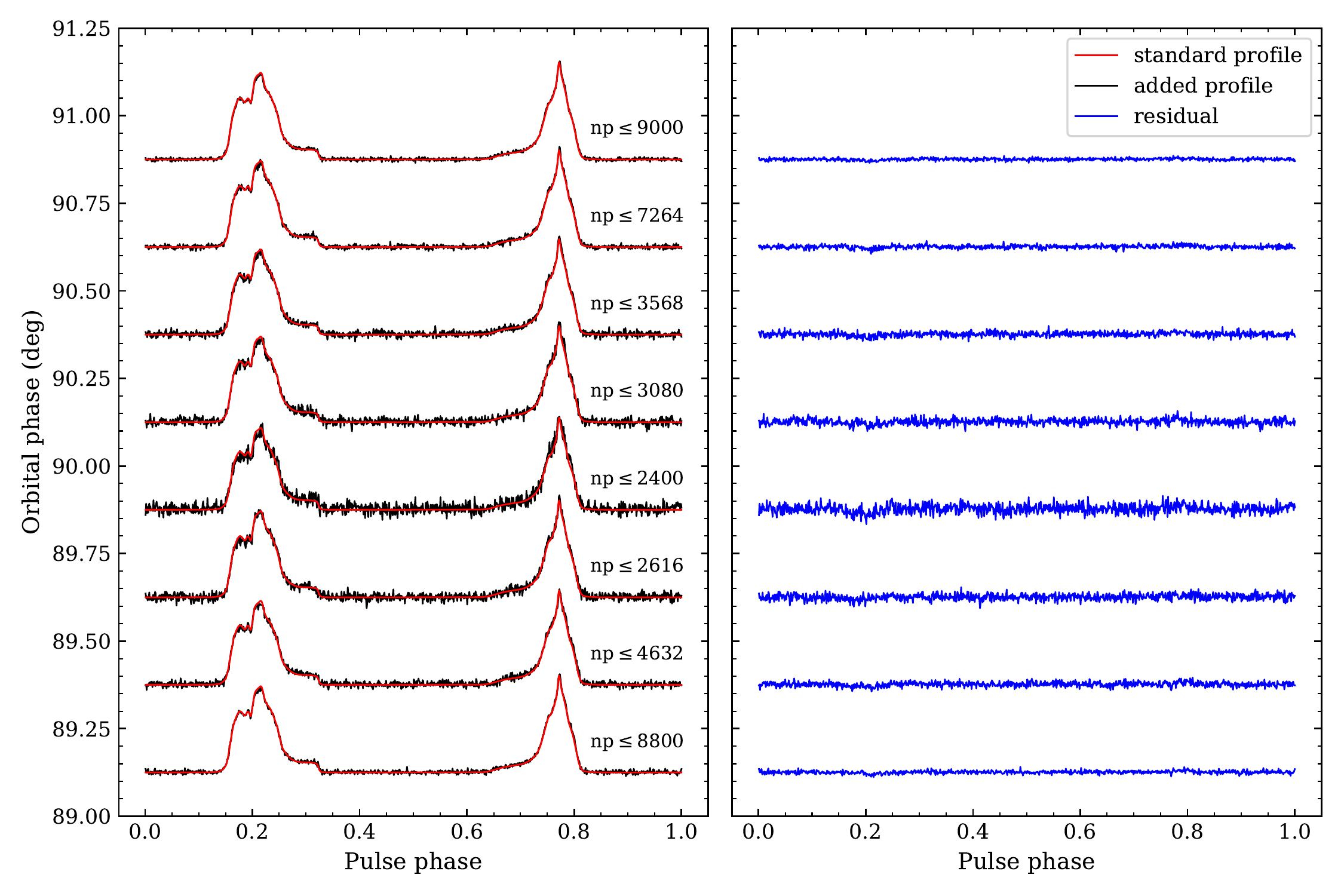}
    \caption{Same as Fig.~\ref{fig:ecl_L} but for the UHF-band data summed from 37 eclipses.}
    \label{fig:ecl_U}
\end{figure*}
\clearpage
}
\subsection{Simulation for latitudinal deflection delay}
\label{sec:lat_time}
To investigate whether the deflection delay caused by latitudinal deflection is measurable from pulsar timing, we perform a simulation based on a simple emission model, which consists of a set of circular cones. Following \citet{DK1995} and \citet{RL2006_rotation}, the latitudinal deflection delay for pulsar A can be written as 
\begin{align}
    \delta_\mathrm{A}^\mathrm{\,latdef} = - \mathcal{D}\, \frac{\cos{i}\sin{(\psi+\delta\psi^\mathrm{ret})}}{(\Lambda_u + \delta \Lambda_u^\mathrm{ret}) \tan{\chi_0}} \,,
    \label{eq:latdefl}
\end{align}
where $\chi_0$ is the angle between the arc connecting the LOS and spin axis and the arc connecting LOS and magnetic axis. Note, Eq.~\eqref{eq:latdefl} is based on the approximation of \citet{DK1995} for the deflection angle, and therefore assumes that the impact parameter is (sufficiently) large compared to the Einstein radius. While this is sufficient, at least until full SKA becomes operational, we present an improved description further below in Section~\ref{sec:SKAlensing}.

We include this deflection time delay in our test model assuming $\chi_0=45\si\degree$~\footnote{Note that $\chi_0$ is not a constant, but our purpose here is to get a feeling for the measurability of the effect in timing rather than a proper account for the effect, which requires knowledge of the latitudinal variation in the emission pattern of the pulsar. Furthermore, as discussed at the beginning of Section~\ref{sec:profile}, the beam geometry adopted by \citet{RL2006_rotation} is not the correct one anyway.}, i.e. a relatively large latitudinal deflection delay, and scale it with a factor $q^\mathrm{\,latdef}$. We simulate high precision TOAs using this test model and fit for the scaling factor. The pre-fit residuals show an advance signature with an amplitude of $-2.8~\mu$s and symmetric to $\psi=90\si\degree$, which is in a similar shape to Shapiro delay but with opposite sign and smaller amplitude. However, after fitting for Shapiro parameters, the signature gets mostly absorbed and leaves residuals below 42~ns at superior conjunction.

\subsection{Prospects of lensing measurement}
\label{sec:lensing}
\begin{figure}[t]
    \centering
    \includegraphics[width=0.9\columnwidth]{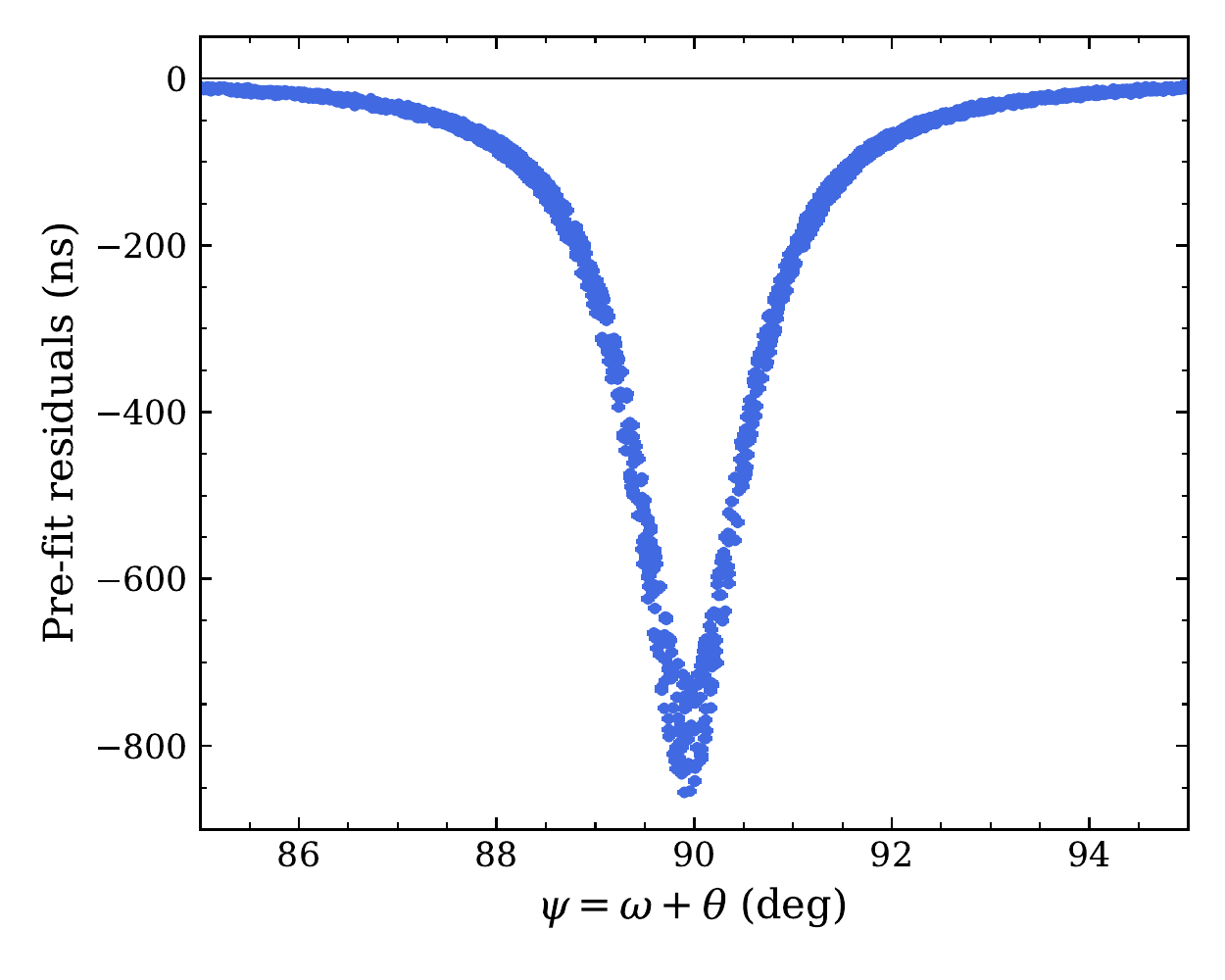}
    \vspace{-5pt}
    \caption{Lensing simulation: pre-fit residual plotted against orbital phase $\psi$. Data displayed here are centred on $\omega=180\si\degree$ and span a decade. The scattering at the lower end of the curve is due to the precession of periastron.}
    \label{fig:len_pre}
\end{figure}
\begin{figure}[t]
    \centering
    \includegraphics[width=0.9\columnwidth]{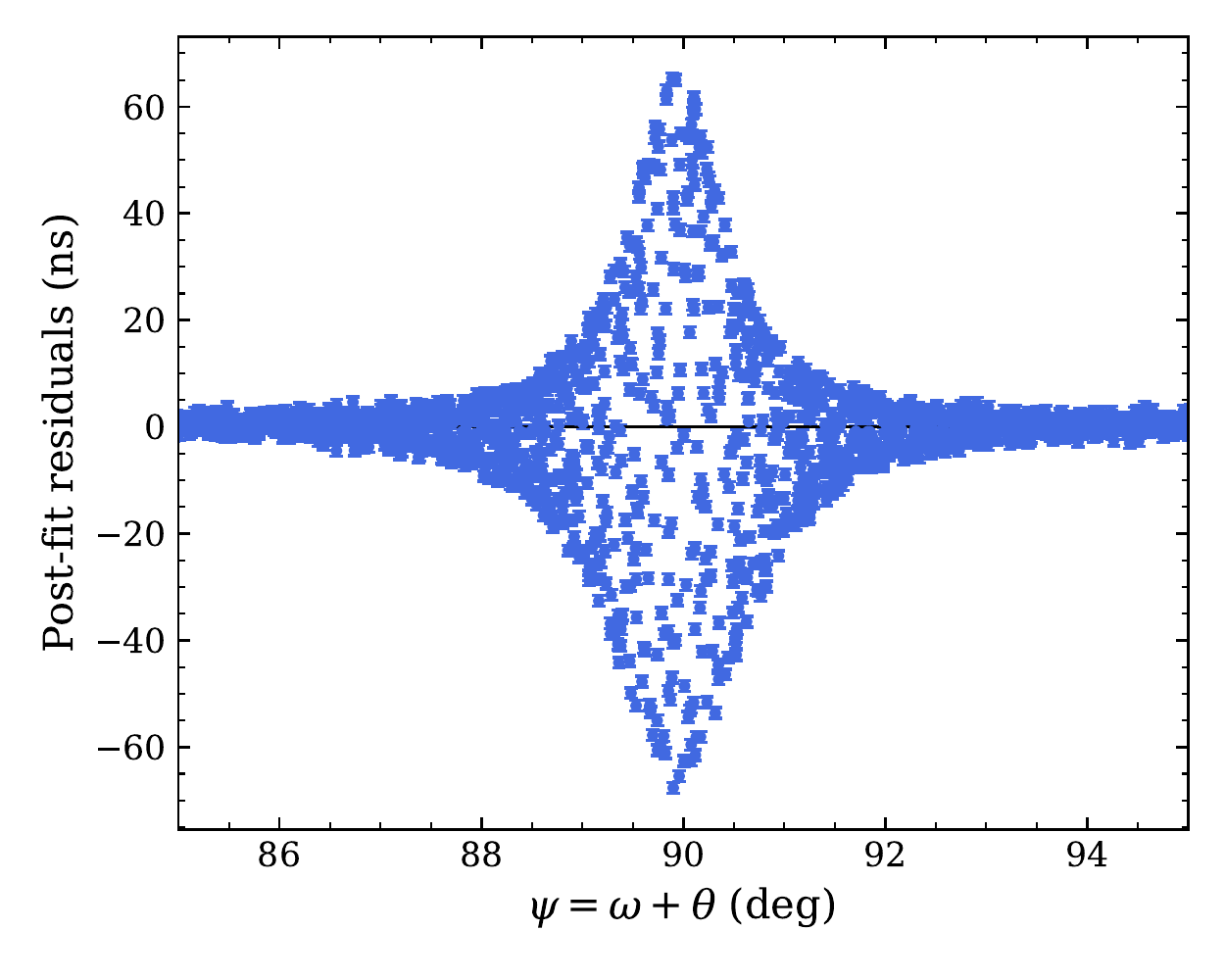}
    \vspace{-5pt}
    \caption{Lensing simulation: post-fit residual plotted against orbital phase $\psi$.}
    \label{fig:len_post}
\end{figure}
\begin{figure}[t]
    \centering
    \includegraphics[width=0.95\columnwidth]{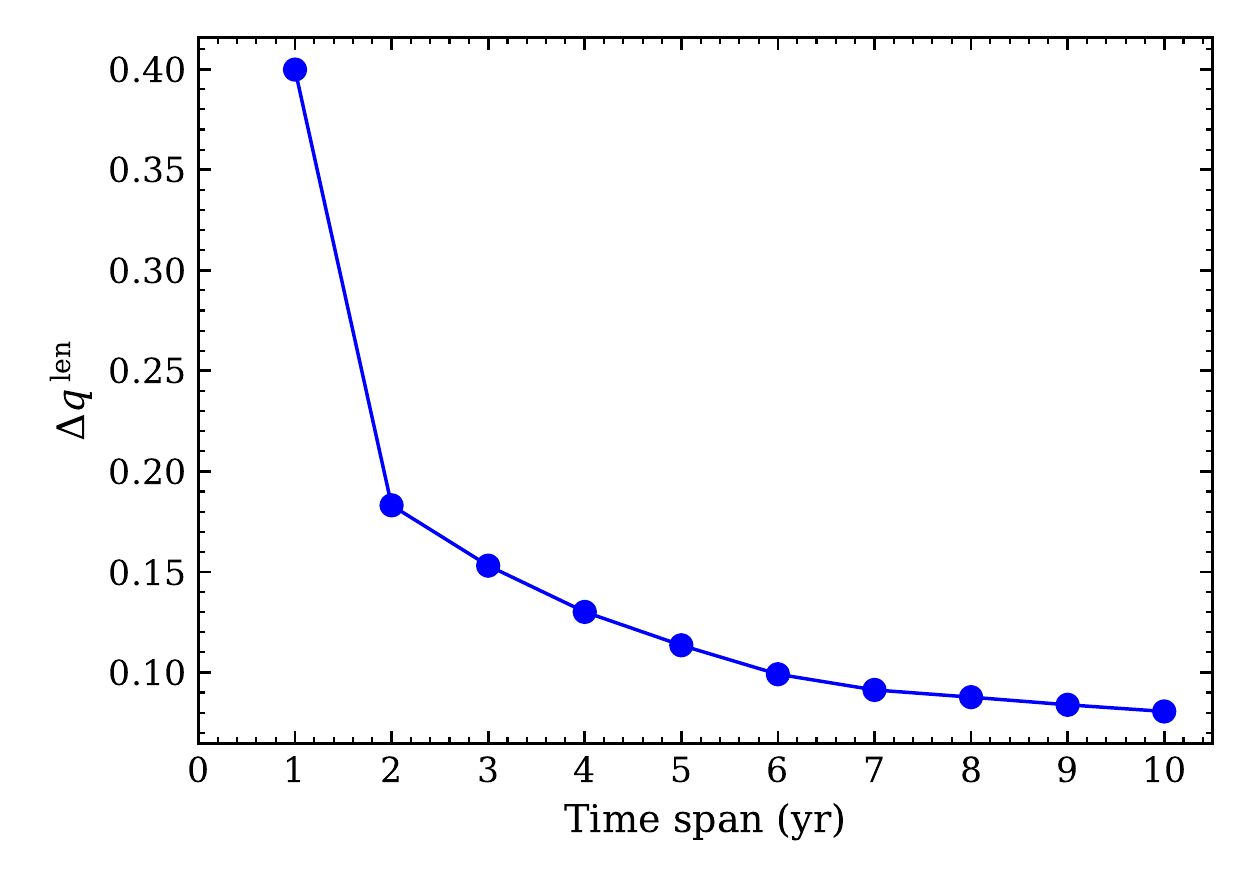}
    \caption{Uncertainty of factor $q^\mathrm{\,len}$ as a function of time span for the simulated data assumed to be 10 times better than the SKA 1.}
    \label{fig:lensing_SKA}
\end{figure}

Even though the retardation and deflection effects can be tested separately while keeping the other one fixed as shown in Eqs.~\eqref{eq:q_def} and \eqref{eq:q_ret}, measuring the lensing correction to Shapiro delay $\delta\Lambda_u^\mathrm{len}$ independently is challenging. As already pointed out by \citet{Kramer+2021DP}, this effect is difficult to observe because of its strong covariance with $s$, or equivalently $z_s$. Our simulation also confirms that the lensing signature can be mostly absorbed by $z_s$ in timing due to its symmetry with respect to conjunction. For demonstration purposes, we simulate 1~ns TOAs which model all NLO signal propagation contributions, then keep the retardation and deflection delay fixed in the model and fit for the scaling factor attached to the lensing correction, $q^\mathrm{\,len}$ (corresponding to $q_\mathrm{NLO}$ in Eq.~\ref{eq:len_nlo}).
Fig.~\ref{fig:len_pre} shows the residuals when lensing correction is not taken into account, leading to a reduced propagation time of about 850~ns as a result of Fermat's principle. Whereas after fitting for $z_s$, the lensing signature gets absorbed and leaves the residuals to be below $70$~ns (Fig.~\ref{fig:len_post}), making a detection with the current timing precision certainly impossible.

To investigate whether lensing can be measured separately in the near future, we simulate TOAs for MeerKAT, MeerKAT extension, and the first phase of the SKA (SKA 1) until 2030 based on similar assumptions made in \citet{Hu2020}. In addition, as the TOA precision reduces significantly due to the intermittent signals during the eclipse (see Fig.~\ref{fig:ecl}), we account for this in our simulations by increasing the uncertainty of these TOAs based on MeerKAT observations. As a simple estimate, we assume GR and perform the simulation using the modified DDGR model with a grid fit to $q^\mathrm{\,len}$.
If lensing is measurable, the value of $q^\mathrm{\,len}$ should be close to $1$. However, it turns out that with the observed and simulated data from 2019 to 2030, the uncertainty of $q^\mathrm{\,len}$ is still larger than 1.

To further push the precision, we assume that in the future an instrument will be available providing a timing accuracy one order of magnitude better than that of the SKA 1 (i.e., 100~ns for an integration time of 30~s), e.g., a future full SKA.
As a rough estimate, here we only consider radiometer noise and ignore any other noise sources, such as jitter noise or scintillation noise. The uncertainty of $q^\mathrm{\,len}$ against the time span of the simulated data is shown in Fig.~\ref{fig:lensing_SKA}. With such precision, one would expect to get a 5-$\sigma$ test of lensing with $\sim 4$~yr of data.

From the simulation, we also obtain an estimated uncertainty for the common factor of NLO contributions $q_\mathrm{NLO}$ in the near future. Assuming no jitter noise, with MeerKAT and the SKA 1, we can expect a 50-$\sigma$ detection by 2030. 

\subsection{Improvements in the timing model for $\lesssim 50$\,ns precision}
\label{sec:SKAlensing}

Equations~(\ref{eq:defl}) and (\ref{eq:latdefl}) are based on the approximation for the signal deflection used by \citet{DK1995}. As discussed in detail in \citet{Kramer+2021DP}, this is still sufficient for the analysis of current timing data. For that reason, the analysis in this paper is still based on \citet{DK1995}, which (including corrections for retardation) is already part of the \textsc{tempo} distribution. However, in the future we can expect to obtain a timing precision of better than $\sim 50$ ns near conjunction ($\pm 1^\circ$), so that an improved treatment of the deflection is required. In a series of papers, Rafikov and Lai have used the standard lensing equation to treat the signal propagation in the Double Pulsar near conjunction \citep{LR2005,RL2006_lensing,RL2006_rotation}. This allowed them to drop the assumption that the impact parameter is much larger than the Einstein radius. The standard lensing equation, however, is based on the assumption of small angles (see e.g. \citealt{sef_book}). Therefore, strictly speaking, Lai and Rafikov's calculations are only valid near conjunction. Similar to calculations of \cite{Shapiro_1967} and \citet{Ward_1970}, \citet{Wucknitz2008} studied the deflection of photons in the gravitational field of a ``point mass'' for general lensing scenarios not limited to regions close to the optical axis. Based on these results, we give an analytic expression for the signal deflection that is valid for the whole orbit, recovers the calculations by Lai and Rafikov near conjunction, and those of \citet{DK1995} if the impact parameter is large compared to the Einstein radius. 

\begin{figure}
    \centering
    \includegraphics[width=\columnwidth]{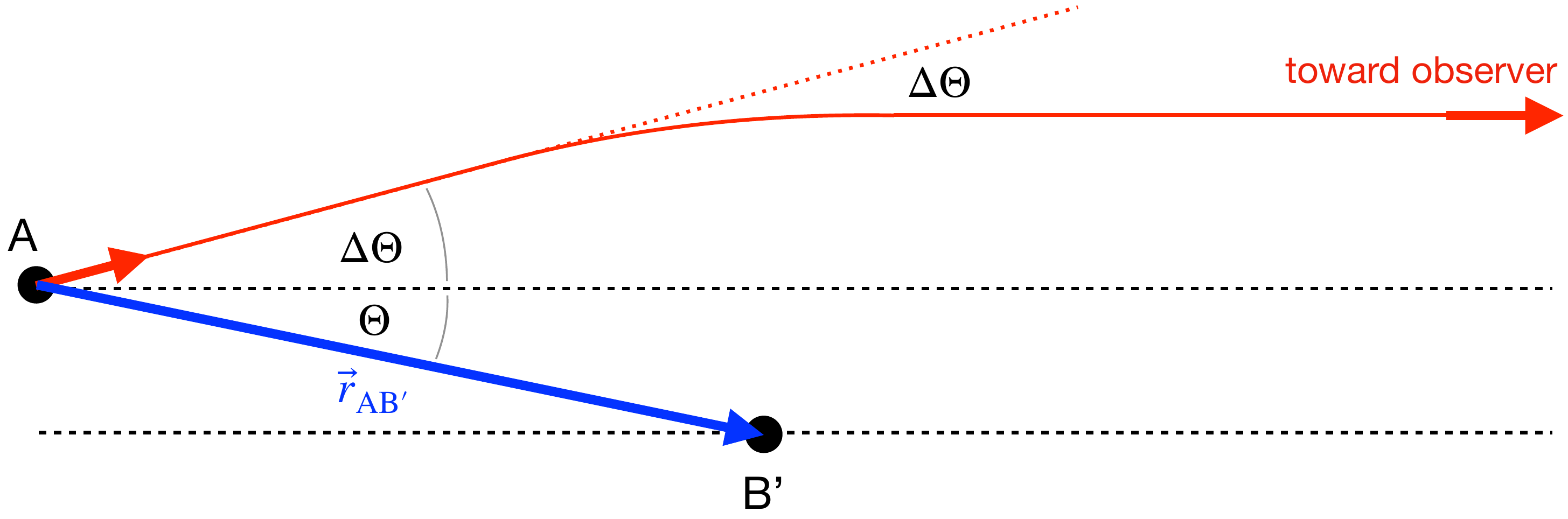}
    \caption{Schematic picture of the lensing geometry as used in Section~\ref{sec:SKAlensing}. B' denotes the retardation corrected position of B (cf.\ \citet{KK1992,KS1999}). In principle there is a second photon path towards the observer (below B'). However, for the Double Pulsar this signal is not only significantly weaker, the path also comes so close to pulsar B that the photons are absorbed by the plasma-filled magnetosphere of B (cf.\ \citealt{LR2005,RL2006_rotation}).}
    \label{fig:lensing_geometry}
\end{figure}

In the following $\Theta$ is the angle between $\vec{r}_\mathrm{AB'}$ and the direction towards the observer, where $\vec{r}_\mathrm{AB'}$ denotes the vector from position of pulsar A at emission to the (retardation corrected) position of pulsar B (the underlying geometry for our calculations is illustrated in Fig.~\ref{fig:lensing_geometry}). The deflection $\Delta\Theta$ of A's radio signal by pulsar B can be obtained from Eq.~(24) in \citet{Wucknitz2008}, with the replacements $\alpha \rightarrow \Delta\Theta$, $\theta \rightarrow \Theta + \Delta\Theta$, and $m \rightarrow \alpha_\mathrm{E}^2$,
where the quantity $\alpha_\mathrm{E}$ is the angle corresponding to the Einstein radius, and is given by
\begin{align}
    \alpha_\mathrm{E} =
    \frac{2}{c}\sqrt{\frac{Gm_\mathrm{B}}{|\vec{r}_\mathrm{AB'}|}}
    \ll 1\,.
\end{align}
This is the maximum value $\Delta\Theta$ can assume. The distance $D_\mathrm{d}$ in \citet{Wucknitz2008} corresponds to $|\vec{r}_\mathrm{AB'}|$.\footnote{Note, the $D_\mathrm{d}$ in Eq.~(24) of \citet{Wucknitz2008} is a typo and should not be there since it is already part of the definition of $m$.} Consequently, one obtains
\begin{align}
    \Delta\Theta \sin(\Theta + \Delta\Theta) 
    - \frac{\alpha_\mathrm{E}^2}{2}\left[1 + \cos(\Theta + \Delta\Theta)\right] = 0 \,.
\end{align}
Since $\Delta\Theta \le \alpha_\mathrm{E} \ll 1$ for all angles $\Theta$, we can expand the equation above in $\Delta\Theta$ while keeping terms only up to order $\alpha_\mathrm{E}^2$. This leads to a quadratic equation 
\begin{align}
    \Delta\Theta^2 + \Delta\Theta\,\sin\Theta
    - \frac{\alpha_\mathrm{E}^2}{2}(1 + \cos\Theta) \simeq 0 \,.
\end{align}
which has, under the assumptions made, one solution:
\begin{align}
    \Delta\Theta \simeq \frac{1}{2}\left(\sqrt{\sin^2\Theta + 2\alpha_\mathrm{E}^2(1 + \cos\Theta)} - \sin\Theta \right) \,.
    \label{eq:DelTheta}
\end{align}
For $\Theta \ll 1$ this agrees with the standard lensing equation (see e.g.\ \citealt{sef_book}).

The angle $\Theta \in [0,\pi]$ needs to be computed from the retardation-corrected orbital phase via $\cos\Theta = \sin i \sin(\psi + \delta\psi^\mathrm{ret})$. The longitudinal and latitudinal deflection delay are given by
\begin{align}
    \delta_\mathrm{A}^\mathrm{\,londef}    &= 
        \frac{\Delta\Theta}{2\pi\nu} \, 
        \frac{\cos(\psi + \delta\psi^\mathrm{ret})}{\sin\Theta\sin i} \,, \\
    \delta_\mathrm{A}^\mathrm{\,latdef} &= 
        -\frac{\Delta\Theta}{2\pi\nu} \,
         \frac{\sin(\psi + \delta\psi^\mathrm{ret})}{\sin\Theta\tan i\tan\chi_0} \,,
\end{align}
respectively (cf.\ Eqs.~(10) and (24) in \citet{RL2006_rotation}, with $\zeta = \pi - i$ and $\eta = -\pi/2$ spin of A aligned with orbital angular momentum). If $\Theta$ is much larger than $\alpha_\mathrm{E}$ one has $\Delta\Theta \simeq \alpha_\mathrm{E}^2(1 + \cos\Theta)/(2\sin\Theta)$. This corresponds to the approximation of \citet{DK1995} for the deflection angle, and one recovers Eqs.~(\ref{eq:defl}) and (\ref{eq:latdefl}).

\section{Discussion}
\label{sec:dis}
In this paper, we have presented results from 3-yr timing observations of the Double Pulsar using the MeerKAT telescope, with a specific focus on studying higher-order signal propagation effects in the gravitational field of the companion. In order to minimise the effects from profile evolution and DM variation, we used frequency-dependent 2D templates to generate TOAs and a DM model to correct dispersive delay in TOAs.

Thanks to its high inclination and orbital compactness, the Double Pulsar is a unique pulsar system for testing NLO signal propagation effects in strong fields. The significantly increased precision offered by MeerKAT permits an independent verification of NLO signal propagation effects and has already surpassed the 16-yr result with only $\sim$3\,yr of data. In our analysis, the Shapiro shape parameter $s$ has improved by 2 times compared to the previous result \citep{Kramer+2021DP}, which also leads to a better mass measurement. The Shapiro range parameter $r$ agrees with GR at $5.3\times 10^{-3}$ ($95\%$ confidence). The precision of the measurement of NLO signal propagation effects $q_\mathrm{NLO}$ has improved by 1.65 times. 
In this work, we investigated the potential profile variation due to latitudinal deflection delay and the possibility of measuring lensing correction to the Shapiro delay, which has never been studied in detail before in pulsar analysis. With the current MeerKAT data, we found little evidence of profile variation at superior conjunction. It could be that the profile variation is not significant at the region we are looking at or our current data are not sensitive enough to identify it. 
We also did simulations on latitudinal deflection delay based on a simple emission model and found it unlikely to be detected because of its correlation with Shapiro delay. 
As for the lensing correction $\delta\Lambda_u^\mathrm{len}$, we found it can be mostly absorbed by the Shapiro shape parameter. Our simulation showed that lensing is unlikely to be measured separately from timing before the full SKA or similarly powerful instruments, and may then be measurable with a few years of timing observations if noises like phase jitter and scintillation do not limit our precision.

However, our analysis also showed that adding certain epochs has a significant impact on the measured Shapiro parameters, but not on $q_\mathrm{NLO}$. 
This could be due to the fact that the phase predicted using the \texttt{polyco} scheme is in particular worse at the superior conjunction, which caused the discrepancies in the Shapiro parameters. Comparison of \texttt{polyco} with different TSPAN values showed residuals oscillating near the superior conjunction, and we may already be limited by the precision of \texttt{polyco} scheme. Of course, there may exist other unknown systematic errors in the data.

To support our timing analysis and study of profile variations at eclipse due to latitudinal deflection, we also checked profiles from all observations. We found variations in the total profile from epoch to epoch.
The differences in the profiles are more prominent at lower frequencies and broadband. Our simulation suggested that these profile variations are likely to be associated with DM variation and scintillation. Even though we have sub-banded data into 16/32 frequency bands and used 2D templates, profile variations may still have an impact on timing. The study of profile variations will be continued in a subsequent work to improve the constraint on the geometry of the system.

Moreover, although not discussed in the paper, we found evidence of red noise in the spectrum of timing residuals with an amplitude two orders of magnitude larger than for typical millisecond pulsars. If not taken into account, it may strongly affect astrometric parameters, as well as influence binary parameters, according to our simulations.
This makes it more difficult to combine the 3-yr MeerKAT data set and the 16-yr data set. Given that the timing precision of the former significantly outperforms the latter, the weighting of MeerKAT data already exceeds the 16-yr data and can dominate noise modelling. For the purpose of this paper, we did not include 16-yr data because of their minor contribution to the $q_\mathrm{NLO}$ measurement ($\sim$10\% improvement). But for studying secular relativistic effects, an appropriate noise modelling may be required to combine these data. We investigate this in further ongoing studies.

{\renewcommand{\arraystretch}{1.2}
\begin{table}[t]
\caption{Comparison of the MeerKAT timing precision $\sigma_\mathrm{RMS}$ assumed in \citet{Hu2020} and from real observations with L-band and UHF-band receivers, scaled to 5-min integration time over the full bandwidth.}
\centering
\begin{tabular}{llc}
\hline \hline
Telescope / receiver & Reference & $\sigma_\mathrm{RMS}$ ($\mu$s) \\ \hline
MeerKAT L band & \citet{Hu2020}   & 1.06 \\
MeerKAT L band & this work  & 0.90\\
MeerKAT UHF band & this work \quad\quad & 0.55\\
\hline
\end{tabular}
\label{tab:compare}
\end{table}
}

In the future, continuing observations with MeerKAT and the SKA will further improve the precision of tests on signal propagation effects, and a 50-$\sigma$ detection of $q_\mathrm{NLO}$ can be expected by 2030. 
For that, we have also provided an improved analytical description of the signal propagation in the Double Pulsar. 
Furthermore, as demonstrated by \citet{Hu2020}, the precision of secular relativistic effects will also be greatly improved and will eventually allow the measurement of the MOI of pulsar A and the NLO gravitational wave damping in the near future. 
The timing precision of the MeerKAT data used in this work is even better than that assumed in \citet{Hu2020}, which is based on early L-band data from MeerKAT (See Table~\ref{tab:compare}). 
This makes their predictions conservative and we are likely to achieve even better measurements with future observations.

\begin{acknowledgements}
We acknowledge Kuo Liu for helpful discussions on data processing and analysis, and Olaf Wucknitz for carefully reading the manuscript and discussions on gravitational lensing which were particularly helpful in Section~\ref{sec:SKAlensing}.  The MeerKAT telescope is operated by the South African Radio Astronomy Observatory, which is a facility of the National Research Foundation, an agency of the Department of Science and Innovation. MeerTime data is housed on the OzSTAR supercomputer at Swinburne University of Technology. 
HH is a member of the International Max Planck Research School for Astronomy and Astrophysics at the Universities of Bonn and Cologne. This work is supported by the Max-Planck Society as part of the ``LEGACY'' collaboration with the Chinese Academy of Sciences on low-frequency gravitational wave astronomy. 
Pulsar research at UBC is supported by an NSERC Discovery Grant and by the Canadian Institute for Advanced Research. Part of this work has been funded using resources from the research grant “iPeska” (P.I. Andrea Possenti) funded under the INAF national call Prin-SKA/CTA approved with the Presidential Decree 70/2016. 
\end{acknowledgements}

%
%
\bibliographystyle{aa}
\bibliography{main.bib}

\end{document}